\documentclass[aps,onecolumn,prd,nofootinbib,
superscriptaddress,preprintnumbers,
longbibliography,eqsecnum,notitlepage,english,10pt]{revtex4-1}
\usepackage{amsfonts}
\usepackage{amsmath}
\usepackage{bm}
\usepackage{babel}
\usepackage{blkarray}
\usepackage{booktabs}
\usepackage{dsfont}
\usepackage[export]{adjustbox}% vertical alignment
\usepackage{graphicx}
\usepackage{nicefrac}
\usepackage[dvipsnames]{xcolor}
\usepackage{braket}
\usepackage{mathtools}
\usepackage{multirow} 

\newcommand{\be}{\begin{equation}}
\newcommand{\ee}{\end{equation}}
\usepackage{hyperref}
\hypersetup{
    colorlinks=true,
    urlcolor=blue,
    linkcolor = blue,
    urlcolor  = blue,
    citecolor = blue,
    anchorcolor = blue,
    pdftitle={ccc},
    pdfauthor={friends},
    pdfsubject={ccc}}
\usepackage{cleveref} %use \cref{}
    \crefformat{figure}{Fig.~#2#1#3}
    \crefformat{table}{Tab.~#2#1#3}
    \crefformat{equation}{Eq.~(#2#1#3)}
    \crefformat{section}{Sect.~#2#1#3}
\graphicspath{{./plots/}}
\usepackage{soulpos}
\ulposdef{\hlst}{%
\rlap{\textcolor{yellow}{\rule[-.75ex]{\ulwidth}{2.5ex}}}%
\rule[.45ex]{\ulwidth}{.1ex}%
}
\newcommand{\diff}{\mathrm{d}}
\newcommand{\cg}[3]{(#1~#2|#3)}
\newcommand\TT{\rule{0pt}{2.6ex}}       % Top strut
\newcommand\BBB{\rule[-1.6ex]{0pt}{0pt}} % Bottom strut

\usepackage{mleftright,xparse}
\NewDocumentCommand\xDeclarePairedDelimiter{mmm}
{%
	\NewDocumentCommand#1{som}{%
		\IfNoValueTF{##2}
		{\IfBooleanTF{##1}{#2##3#3}{\mleft#2##3\mright#3}}
		{\mathopen{\csname##2\endcsname#2}##3\mathclose{\csname##2\endcsname#3}}%
	}%
}
\xDeclarePairedDelimiter{\abs}{\lvert}{\rvert}

\begin{document}

%%%%%%%%%%%%%%%%%%%%%%%%%%%%%%%%%%%%%%%%%%%%%%%
% \date{\today}

\title{
A unitary coupled-channel three-body amplitude with pions and kaons
}
\author{Yuchuan Feng}
\email{fengyuchuan@gwmail.gwu.edu}
\affiliation{The George Washington University, Washington, DC 20052, USA}
\author{Fernando Gil}
\email{fernando.gil@ific.uv.es}
\affiliation{IFIC (CSIC-UV,
Catedrátic José Beltrán Martinez 2, Valencia, Spain)}
\author{Michael~D\"oring}
\email{doring@gwu.edu}
\affiliation{The George Washington University, Washington, DC 20052, USA}
\affiliation{Theory Center, Thomas Jefferson National Accelerator Facility, Newport News, VA 23606, USA}
\author{Raquel Molina}
\email{raquel.molina@ific.uv.es}
\affiliation{IFIC (CSIC-UV,
Catedrátic José Beltrán Martinez 2, Valencia, Spain)}
%%%
\author{Maxim~Mai}
%\email{maximmai@gwu.edu}
\affiliation{The George Washington University, Washington, DC 20052, USA}
\affiliation{Albert Einstein Center for Fundamental Physics, Institute for Theoretical Physics, University of Bern, Sidlerstrasse 5, 3012 Bern, Switzerland}
\affiliation{Helmholtz-Institut für Strahlen- und Kernphysik (Theorie) and
Bethe Center for Theoretical Physics, Universität Bonn, 53115 Bonn, Germany}
\author{Vanamali~Shastry}
\affiliation{Department of Physics, Indiana University, Bloomington, IN 47405, USA}
\affiliation{Center for Exploration of Energy and Matter, Indiana University, Bloomington, IN 47403, USA}
\author{Adam~Szczepaniak}
\affiliation{Department of Physics, Indiana University, Bloomington, IN 47405, USA}
\affiliation{Center for Exploration of Energy and Matter, Indiana University, Bloomington, IN 47403, USA}
\affiliation{Theory Center, Thomas Jefferson National Accelerator Facility, Newport News, VA 23606, USA}

%%%%
%%%%%%%%%%%%%%%%%%%%%%%%%%%%
\preprint{JLAB-THY-24-4107}

%%%%%%%%%%%%%%%%%%%%%%%%%%%%%%%%%%%%%%%%%%%%%%%%%%%%%%%
%%%%%%%%%%%%%%%%%%%%%%%%%%%%%%%%%%%%%%%%%%%%%%%%%%%%%%%
\begin{abstract}
Three-body dynamics above threshold is required for the reliable extraction of many amplitudes and resonances from experiment and lattice QCD. The S-matrix principle of unitarity can be used to construct dynamical coupled-channel approaches in which three particles scatter off each other, re-arranging two-body subsystems by particle exchange. This paper reports the development of a three-body coupled-channel, amplitude including pions and kaons. The unequal-mass amplitude contains two-body S- and P-wave subsystems (``isobars'') of all isospins, $I=0,\,\nicefrac{1}{2},\,1, \nicefrac{3}{2}, \, 2$,  and  it also allows for transitions within a given isobar. The $f_0(500)\, (``\sigma''),\,f_0(980),\,\rho(700), K_0^*(700)\,(``\kappa'')$, and $K^*(892)$ resonances are included, apart from repulsive isobars. Different methods to evaluate the amplitude for physical momenta are discussed. Production amplitudes for $a_1$ quantum numbers are shown as a proof of principle for the numerical implementation.
\end{abstract}
%%%%%%%%%%he resca%%%%%%%%%%%%%%%%%%%%%%%%%%%%%%%%%%%%%%%%%%%%%
%%%%%%%%%%%%%%%%%%%%%%%%%%%%%%%%%%%%%%%%%%%%%%%%%%%%%%%

\maketitle

%\tableofcontents

%%%%%%%%%%%%%%%%%%%%%%%%%%%%%%%%%%%%%%%%%%%%%%%%%%%%%%%
%%%%%%%%%%%%%%%%%%%%%%%%%%%%%%%%%%%%%%%%%%%%%%%%%%%%%%%
\section{Introduction}
%%%%%%%%%%%%%%%%%%%%%%%%%%%%%%%%%%%%%%%%%%%%%%%%%%%%%%%
%%%%%%%%%%%%%%%%%%%%%%%%%%%%%%%%%%%%%%%%%%%%%%%%%%%%%%%
Quantum Chromodynamics (QCD) governs the dynamics of strongly interacting matter. Its quantum states, the ground- and excited-state hadrons, exhibit a complex mass and decay pattern that is difficult to explain directly from QCD. Many theoretical tools have been developed in the past to address this issue, for example Effective Field Theories~\cite{Meissner:1993ah,Holstein:2008zz,Epelbaum:2010nr,Hermansson-Truedsson:2020rtj,Mai:2022eur}, Lattice QCD (LQCD)~\cite{Briceno:2017max,Bulava:2022ovd}, and functional methods~\cite{Eichmann:2016yit, Eichmann:2016hgl, Yin:2019bxe}; for a recent review see Ref.~\cite{Mai:2022eur}. One complication is that the elastic energy window is usually small, making it necessary to consider coupled channels and three or more body dynamics. Indeed, already the first excited state of the nucleon, the so-called Roper-resonance $N^*(1440)1/2^+$, is above the $\pi\pi N$ threshold, decaying substantially into effective three-body channels $f_0(500)N$  and $\pi\Delta$ causing the unusual line shape of that resonance~\cite{Arndt:1995bj, Alvarez-Ruso:2010ayw, Ronchen:2012eg, MartinezTorres:2008kh}. Some resonances like the $a_1(1260)$ meson decay overwhelmingly into three particles~\cite{ParticleDataGroup:2022pth}. The same is expected for the lightest meson with exotic spin-parity quantum numbers, $J^{PC} = 1^{-+}$, that is currently a subject of extensive experimental and theoretical investigations~\cite{COMPASS:2021ogp, Chen:2022asf}, see also  Refs.~\cite{Liu:2024uxn, JPAC:2021rxu} for recent reviews. Three-body effects can also be relevant for the other exotic candidates,  
like the X(3872)~\cite{Baru:2011rs}, the Y(4260)~\cite{Cleven:2013mka}, or the $T_{cc}$(3875)~\cite{Du:2021zzh,Lin:2022wmj}, 
 observed in the charmonium spectrum close to various three-body thresholds. 
Since a few years, it has become possible to calculate three-body systems from LQCD due to two closely intertwined factors. First, increased computational capacities and algorithmic developments have made the implementation of multi-hadron operators possible, see  
%==== Only whose with new Lattice calculations until end of 2023 %
Refs.~\cite{Lang:2016hnn,Kiratidis:2016hda,Liu:2016uzk,Horz:2019rrn,Culver:2019vvu,Fischer:2020jzp,Hansen:2020otl,Alexandru:2020xqf,Blanton:2021llb,NPLQCD:2020ozd,Buhlmann:2021nsb,Mai:2021nul,Garofalo:2022pux,Draper:2023boj} for recent developments. 
Such programs provide a discrete finite-volume energy spectrum, which encodes the full QCD dynamics. Relating this spectrum to the infinite-volume (coupled-channel) amplitude is accomplished by  so-called Quantization Conditions (QCs), see 
%% ==== Developments of the 3bQC ===
%Refs.~\cite{Beane:2007es, Polejaeva:2012ut,Briceno:2012rv,Meissner:2014dea,Hansen:2014eka,Jansen:2015lha,Hansen:2015zga,Hansen:2015zta,Hansen:2016fzj,Guo:2016fgl,Mai:2017bge,Konig:2017krd,Hammer:2017uqm,Hammer:2017kms,Briceno:2017tce,Sharpe:2017jej,Guo:2017crd,Guo:2017ism,Meng:2017jgx,Guo:2018ibd,Guo:2018xbv,Klos:2018sen,Briceno:2018mlh,Briceno:2018aml,Mai:2018djl,Doring:2018xxx,Jackura:2019bmu,Mai:2019fba,Guo:2019hih,Blanton:2019igq,Briceno:2019muc,Romero-Lopez:2019qrt,Pang:2019dfe,Guo:2019ogp,Zhu:2019dho,Pang:2020pkl,Hansen:2020zhy,Guo:2020spn,Guo:2020wbl,Guo:2020ikh,Guo:2020kph,Blanton:2020gha,Blanton:2020gmf,Muller:2020vtt,Muller:2020wjo,Jackura:2020bsk,Brett:2021wyd,Muller:2021uur,Grabowska:2021xkp,Hansen:2021ofl,Blanton:2021mih,Blanton:2021eyf,Jackura:2022gib,Muller:2022oyw,Pang:2022nim, Jackura:2023qtp,Raposo:2023oru,Pang:2023jri,Bubna:2023oxo} and  
Refs.~\cite{Hansen:2019nir,Mai:2021lwb} for dedicated reviews in the three-hadron sector and Ref.~\cite{Mai:2021nul} for a first application to a three-body resonance using two channels. Recent finite-volume extensions to be highlighted are the coupled-channel, unequal mass cases of Ref.~\cite{Draper:2023boj} and Ref.~\cite{Draper:2024qeh} ($\eta\pi\pi$ and $K\bar K\pi$ channels), as well the solution of integral equations in this context~\cite{Hansen:2020otl}. A common feature of the pertinent amplitudes is the need to explicitly incorporate particle exchange to all orders in rescattering of the amplitude as dictated by unitarity, see, e.g., Refs.~\cite{Mai:2017bge, Mai:2021lwb}. 

In experimental analyses of three-body states, the traditional ``isobar model'' employs lineshapes (isobars) for the final 2-body correlation in the decay, without explicit three-body  rescattering effects, e.g., at Crystal Barrel~\cite{CrystalBarrel:2019zqh} (includes coupled channels), BESIII~\cite{BESIII:2023qgj}, Belle~\cite{Rabusov:2022woa}, and LHCb~\cite{LHCb:2022lja}. 
Recently the COMPASS collaboration significantly advanced the analysis of the three-pion spectrum~\cite{Ketzer:2019wmd, COMPASS:2018uzl}  by performing independent partial wave decomposition in bins of the two-pion subsystems, i.e., without introducing isobars of specific mass in the two pion subsystem~\cite{Krinner:2017dba}. 

In this paper, we concentrate on the development of a unitary (infinite-volume) amplitudes with explicit rescattering and coupled channels, that can be extended for the analysis of experimental data, but also for QCs in future work. This work builds on the two-channel approaches of Refs.~\cite{Sadasivan:2021emk, Mai:2021nul} in which, for the first time, a three-body resonance pole was extracted from data with a unitary amplitude. See Ref.~\cite{JPAC:2018zwp} for an extraction with approximate unitarity and Ref.~\cite{Molina:2021awn} for first steps to enlarge the channel space. Here, we are interested in the case of strangeness zero and allow for pions and kaons in the amplitude. In future work, $\eta$ mesons and other hadrons will be included. Here, we formulate nine channels that not only contain canonical S- and P-wave resonances in the two-body subsystems --the $f_0(500)$ ``$\sigma$'', $f_0(980)$, $\rho(770)$, $K_0^*(700)$ ``$\kappa$'', $K^*(892)$-- but also repulsive $\pi\pi$ and $\pi K$ subsystems at maximal isospin. Amplitudes are formulated for all total isospins $I=0,\dots,3$ with zero overall strangeness. To discuss numerical aspects, we solve the amplitude for the $a_1$ quantum numbers in a minimal fashion, i.e., with exchange processes,  but without the aim to describe phenomenology like the $a_1(1260)$ resonance or the $a_1(1420)$ triangle singularity (TS) discussed below. 

In experimental analysis, the description of three-body final states with unitary isobar amplitudes has made substantial progress recently. 
See Refs.~\cite{Mai:2017vot, Jackura:2018xnx, Jackura:2019bmu, Albaladejo:2019huw, Mikhasenko:2019vhk, Jackura:2020bsk, MartinezTorres:2020hus, Zhang:2021hcl, Jackura:2023qtp} for formal developments. In close conceptual connection to the present effort, in Ref.~\cite{Nakamura:2023hbt} a three-body unitary coupled-channel framework was used to analyze $K_SK_S\pi^0$ Dalitz plot pseudodata for the $J^{PC}=0^{-+}$ amplitude in the $J/\psi\to \gamma K_SK_S\pi^0$ decay related to the controversial $\eta(1405/1475)$, extracting also the associated resonance poles. 
That framework is conceptually related to pioneering work by the ANL-Osaka group~\cite{Kamano:2011ih, Nakamura:2012xx}. See also Refs.~\cite{Nakamura:2023obk, Zhang:2024dth} for other unitary three-body coupled-channel applications. Three-body rescattering effects can also be quantified using Khuri-Treimann (KT) equations. While this is a separate topic, we mention here Ref.~\cite{Stamen:2022eda} in which KT and Omn\`es-functions for the two-body part were used to study  $\pi\rho$ rescattering in close relation to the physical system studied here. 

As mentioned before, three-body effects can be enhanced 
 by a kinematical effect known as the singularity (TS).
  It 
%, which became popular in the sixties~\cite{Karplus:1958zz,Landau:1959fi,Booth:1961zz,Anisovich:1964ikk}, 
 happens in a decay of a resonance to an intermediate   resonance  and a spectator. In a specific kinematic region one of the decay products of the intermediate resonance can "catch up" with the spectator and produce a bump in the invariant  mass spectrum of these two particles. 
 %Subsequently, particle 2 decays into 4+5, and one of %those fuses with 3, in such a way that all particles %inside of the loop diagram can be placed on-shell, %leading to an enhanced cross section.
%, they are collinear, and the process can occur at the classical level, as demanded by the Coleman-Norton theorem~\cite{Coleman:1965xm}. 
See also Refs.~\cite{Liu:2015taa,Bayar:2016ftu} for the mathematical treatment. The $a_1(1420)$, was at first
 claimed to be a new resonance while posterior analyses suggested that a TS effect can take place in the decay chain $a_1(1260) \to K^* \bar K \to K\bar K \pi \to f_0 \pi \to 3\pi$ resulting in a peak in the $3\pi$ mass $~0.2\mbox{ GeV}$ above the nominal $a_1$ mass~\cite{Mikhasenko:2015oxp, Guo:2019twa, COMPASS:2020yhb}, see also Refs.~\cite{Bayar:2016ftu,Debastiani:2018xoi, Guo:2019twa, Isken:2023xfo}  for general discussions. It is also well-known that triangle singularities play an important role in two nucleon fusion reactions like $pp\to \pi^+d$, which can lead to the generation of the so-called dibaryon peak~\cite{Molina:2021bwp}.  
%See Refs.~\cite{Dai:2018rra, Liang:2019jtr, Jing:2019cbw} for further examples of TS. The exact position of the TS can also be used to determine the mass of an exotic state close to a three-body threshold when the decay channel of the exotic particle is mediated by a TS~\cite{Guo:2019qcn,Molina:2020kyu}.

Another analysis step requires the analytic continuation of the amplitude to the resonance poles to determine their position and residues, corresponding to masses and branching ratios. For three-body states, this can be achieved, for example, by contour deformation~\cite{Doring:2009yv, Suzuki:2009nj, Sadasivan:2021emk, Dawid:2023jrj,Dawid:2023kxu}. Mesonic resonance poles in three-body amplitudes were only extracted recently~\cite{Sadasivan:2021emk, Mai:2021nul,Nakamura:2023obk}, while the extraction of exited baryon resonance poles including three-body channels has been carried out by various groups since a long time~\cite{Doring:2009yv, Suzuki:2009nj}. 
For the present amplitude, one can directly follow Ref.~\cite{Sadasivan:2021emk} which should be consulted for specific details. 

This paper is organized as follows. In Sec.~\ref{sec:introsimple} the kinematics and general ideas for the construction of the three-body amplitude are introduced. In Secs.~\ref{sec:channels} and \ref{sec:transitions} the selection of coupled channels is discussed and formalized with respect to their isospin and $G$-parity structure. The isobar-spectator interaction is matched to experimental input from coupled-channel two-body scattering in Sec.~\ref{sec:props}. The transformation from the plane-wave, helicity formulation to the basis with definite total angular momentum $J$, spin $S$, and orbital angular momentum (JLS basis) is described in Sec.~\ref{sec:jls}. The remainder of Sec.~\ref{sec:formalism} contains a technical discussion on how to solve the JLS-projected integral equations for the scattering matrix, with numerical results shown in Sec.~\ref{sec:results}. 
Sec.~\ref{sec: conclusion} concludes the paper and briefly discusses future applications and extensions of the model to semileptonic $\tau$ decays and related reactions, comparison to traditional isobar model and the study of triangle singularities. In appendix.~\ref{sec:lagrangians}, the calculation of isospin coefficients in Sec.~\ref{sec:transitions}
is compared to a Lagrangian formalism for double checking.

%%%%%%%%%%%%%%%%%%%%%%%%%%%%%%%%%%%%%%%%%%%%%%%%%%%%%%%
%%%%%%%%%%%%%%%%%%%%%%%%%%%%%%%%%%%%%%%%%%%%%%%%%%%%%%%
\section{Formalism}
\label{sec:formalism}
%%%%%%%%%%%%%%%%%%%%%%%%%%%%%%%%%%%%%%%%%%%%%%%%%%%%%%%
%%%%%%%%%%%%%%%%%%%%%%%%%%%%%%%%%%%%%%%%%%%%%%%%%%%%%%%

%%%%%%%%%%%%%%%%%%%%%%%%%%%%%%%%%%%%%%%%%%%%%%%%%%%%%%%
\subsection{Coupled-channel three-body amplitude}
\label{sec:introsimple}
\begin{figure}[htb]
\begin{center}
\includegraphics[width=0.8\textwidth]{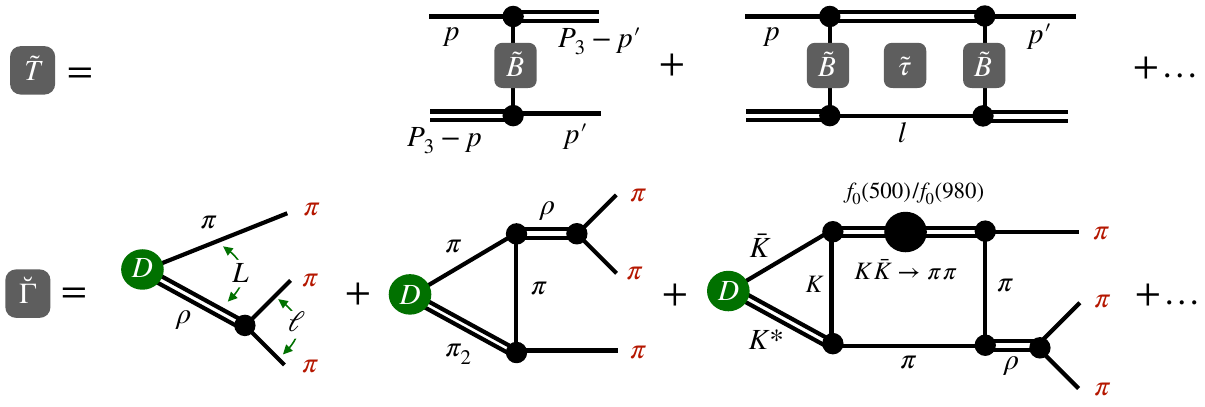}
\end{center}
\caption{Upper row: The rescattering series $\tilde T$ of isobars (double lines) and spectators (single lines) interacting through particle exchanges $\tilde B$, followed by propagation $\tilde \tau$. The amplitude $\tilde T(s,p',p)$ depends on total Mandelstam $s=P_3^2$ and incoming (outgoing) spectator momentum $p$ ($p'$), while the isobar propagation depends on $\sigma=(P_3-p)^2$, i.e., $\tilde\tau(\sigma(l))$ for the internal propagation shown.
Lower row: The green bubble $D$ is the elementary decay process without coupled-channel effects, defined in Eq.~\eqref{eq:D}.
This series forms the final-state interaction in the production reaction $\breve \Gamma(s,p')$ in the lower row. For $\breve \Gamma(s,p')$,  some coupled channels are indicated that contribute to a three-pion final state, as well as the orbital angular momentum $L$ and isobar spin $\ell$.
}
\label{fig:ccillu}
\end{figure}

%%%%%%%%%%%%%%%%%%%%%%%%%%%%%%%%%%%%%%%%
To describe the interaction of three particles, one can cast it in the interactions between all possible two-particles subsystem, the ``isobars'', each of them accompanied with a ``spectator''.
Such an interaction is shown as $\tilde B$ in Fig.~\ref{fig:ccillu}. 
In notation we follow here Ref.~\cite{Mai:2017vot} that is based on classic works, {\it e.g.} Ref. ~\cite{Aaron:1968aoz}.
While the isobars may or may not contain two-body resonances, they can also contain two-body coupled channels leading to $\pi\pi\leftrightarrow K\bar K$ transitions as indicated to the lower right in the figure. 

The isobar-spectator rescattering series $\tilde T$
can be cast in a Lippmann-Schwinger-like equation with relativistic kinematics~\cite{Mai:2017vot, Mai:2017bge}. For the channel transition $i\to j$ it reads
%%%%%%%%%%%%
\begin{align}
    \tilde T_{ji}(s,{\bm p}',\bm{p})=
    \tilde B_{ji}(s,{\bm p}',\bm{p})+
    \tilde C_{ji}(s,{\bm p}',\bm{p})+
    \int\frac{\mathrm{d}^3l}{(2\pi)^3\,2E_{l,k}}
    \left(
    \tilde B_{jk}(s,{\bm p}',{\bm  l})+\tilde C_{jk}(s,{\bm p}',{\bm  l })
    \right)\,
    \tilde \tau_k(\sigma_l) \, 
    \tilde T_{ki}(s,{\bm l},{\bm{p}}) \ ,
    \label{eq:T3-integral-equation}
\end{align}
%%%%%%%%%%%%
with the total Mandelstam $s$ and the isobar Mandelstam $\sigma$, as momenta as indicated in the figure, and
\begin{align}
    s=P_3^2,\quad \sigma=(P_3-p)^2=s+m_k^2-2\sqrt{s}E_{l,k} 
    \ ,\quad\text{where}\quad E_{l,k}=\sqrt{m_k^2+l^2} \ ,
\end{align}
 for the total four-momentum $P_3$ and spectator mass $m_k$ in channel $k$ (repeating indices like $k$ imply summation). Here, we work in the three-body center of mass, $\bm{P}_3=\bm{0}$. 
Note that Eq.~\eqref{eq:T3-integral-equation} is obtained from a (covariant) Bethe-Salpeter equation
by maintaining only the positive-energy component of the spectator and putting it onshell, ${\tilde\tau(\sigma(l))=2\pi\delta(l^2-m_k^2)\theta(l^0)\,S}$, where $S$ is the generic isobar-spectator propagation~\cite{Mai:2017vot}. This is part of a matching process in which three-body unitarity is implemented in the amplitude $\tilde T$, requiring $\tilde B$ to develop an imaginary part consistent with the imaginary part of $\tilde\tau^{-1}$; this is also the reason why $\tilde B$ is written separately from other three-body interactions $\tilde C$ that respect unitarity as long as they are real~\cite{Mai:2017vot} in the physical region.

The scattering equation~\eqref{eq:T3-integral-equation} is correct for both distinguishable and  indistinguishable particles. The reason is that symmetry factors are absorbed in the definition of $\tilde\tau$, see the discussion in Sec.~\ref{sec:tau}. 
In this work, the tilde notation is used to distinguish $\tilde B$ and $\tilde\tau$ from their counterparts in previous publications, see, e.g., Ref.~\cite{Sadasivan:2021emk}. The reason is that, here, isospin factors are distributed differently between these quantities, as explained in the following.

The production process $\breve\Gamma$ shown in in the second row of Fig.~\ref{fig:ccillu} contains the unitary final-state interaction $\tilde T$. It also obeys a integral equation like $\tilde T$, as discussed in Sec.~\ref{sec:direct}. We mention here only that the elementary production process, $D$, has to be real and channel-dependent like $\tilde C$ and, like that term, it can depend on $s$ and spectator momentum, and both terms together can be used to describe production reactions, like semileptonic $\tau$ decays~\cite{Sadasivan:2020syi}.

%%%%%%%%%%%%%%%%%%%%%%%%%%%%%%%%%%%

\subsection{Channel space}
\label{sec:channels}
The helicity basis allows to express states in terms of:
\begin{enumerate}
\item
The isobar helicity (the spectator is, in the present case, always a (spinless) pseudoscalar meson). For cases in which both isobar and spectator have spin, see, e.g., Ref.~\cite{Ronchen:2012eg}.
\item 
The particle type of the spectator and the quantum numbers of the isobar. For a given final three-body state, one has to sum over all combinations leading to the given overall quantum number. This is usually where model dependence comes into the formalism through a truncation of the considered channels to keep the implementation technically manageable.  
\item
The particle type in a given isobar. This distinction is required when the isobar itself is a two-channel system. In a system with total strangeness zero, both the isoscalar and isovector channels could be coupled $\pi\pi\leftrightarrow K\bar K$ systems. We neglect here the small inelasticity of the $\rho$ channel into $K\bar K$, but we take the $\pi\pi\leftrightarrow K\bar K$ dynamics for the isoscalar channel into account, that is crucial for the $f_0(980)$ resonance which resides in the same partial wave as the $\sigma$ resonance. 

For the vector $K^*$ and scalar $\kappa$ channel, one has a strong $\pi K$ channel but could also have $\eta K$. As this work focuses on pions and kaons, we treat these channels as mostly elastic except that for the $\kappa$ we allow for a small inelasticity from $\eta K$ owing to the way this channel is programmed, see Sec.~\ref{sec:results}. We also ignore the small P-wave isospin $I_I=\nicefrac{3}{2}$ $\pi K$ channel and only consider the S-wave channel, named $K_{\nicefrac{3}{2}}$ in the following.
\end{enumerate}
The resulting channel space is shown in Table~\ref{tab:channels} for the considered quantum numbers of the $a_1(1260)$, i.e., negative $G$-parity $\eta_G=-1$. All channels would contribute to total isospin $I=1$, but only some would contribute to isospins $I=0$, $I=2$, and $I=3$.
%%%%%%%%%%%%%%%%%%%%%%%%%%%%%%%%%%%%%%%%%%%%%%
\begin{table}[tb]
\center
    \begin{tabular}{l|l|l|l|l|l|l|l|l|l}
    \hline\hline 
Isobar $(\ell,I_I)$
&
\multicolumn{2}{c|}{$(1,1)$}
&
\multicolumn{2}{c|}{$(1,\nicefrac{1}{2})$}
&
\multicolumn{2}{c|}{$(0,0)$}
&
$(0,2)$ & $(0,\nicefrac{1}{2})$ & $(0,\nicefrac{3}{2})$
\TT \\
HB basis (11 Ch.)
& 
\multicolumn{2}{c|}{$\pi\rho_{\lambda=\pm 1, 0}$}
& 
\multicolumn{2}{c|}{$KK^*_{\lambda=\pm 1, 0}$}
&$\pi\sigma$ & $\pi (K\bar K)_S$
& $\pi\pi_2$ & $K\kappa$ & $KK_{\nicefrac{3}{2}}$
\\
JLS basis (9 Ch.)
&
$(\pi\rho)_S$ & $(\pi\rho)_D$
&
$(KK^*)_S$ & $(KK^*)_D$
&$(\pi\sigma)_P$ & $(\pi (K\bar K)_S)_P $
& $(\pi\pi_2)_P$ & $(K\kappa)_P$ & $(KK_{\nicefrac{3}{2}})_P$
\BBB\\
\hline
\hline
    \end{tabular}
\caption{Channels for $\eta_G=-1$. The spin $\ell$ and isospin $I_I$ of the isobar is indicated in the first line for clarity. The second line shows the eleven channels in helicity basis (HB). The third line indicates the nine channels in the JLS basis for total angular momentum $J=1$. The outermost subscripts indicate the relative angular momentum $L$ between isobar and spectator. The abbreviation $\pi_2$ stands for the S-wave isospin $I_I=2$ repulsive $\pi\pi$ channel and $K_{3/2}$ for the repulsive S-wave $I_I=\nicefrac{3}{2}$ $\pi K$ channel.}
\label{tab:channels}
\end{table}
%%%%%%%%%%%%%%%%%%%%%%%%%%%%%%%%%%%%%%%%%%%
Note that positive/negative $G$-parity requires the channels containing kaons to be combined according to
\begin{align}
\eta_+=\frac{1}{\sqrt{2}}\left(\bar K_{I_I} K+K_{I_I} \bar K\right) \ , \quad
\eta_-=\frac{1}{\sqrt{2}}\left(\bar K_{I_I} K-K_{I_I}\bar K\right) \ ,
\label{gparity}
\end{align}
where $K_{I_I}$ stands for the S-wave or P-wave isobars in isospin $I_I=\nicefrac{1}{2}$ or $I_I=\nicefrac{3}{2}$. In Table~\ref{tab:channels}, we indicate the combinations $\eta_-$ as $KK^*$, $K\kappa$, or $KK_{\nicefrac{3}{2}}$.

%%%%%%%%%%%%%%%%%%%%%%%%%%%%%%%%%%%%
\subsection{Channel Transitions}
\label{sec:transitions}
\begin{figure}[tb]
\begin{center}
\includegraphics[width=0.35\textwidth]{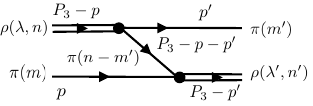}
\end{center}
\caption{The $\pi\rho\to\pi\rho$ transition. The labels $n^{(\prime)},\,m^{(\prime)}$ stand for third components of isospin of the incoming (outgoing) states.}
\label{fig:exlabels}
\end{figure}

The exchange transitions required by unitarity~\cite{Mai:2017vot}, with the labeling of Fig.~\ref{fig:exlabels}, read in the helicity basis (HB):
\begin{align} 
    \tilde B_{ji}(s,\bm{p}',\bm{p})&
    =\frac{(\tilde {\bm I}_F)_{ji}\,
    v_j^*(p,P-p-p')
    v_i(p',P-p-p')}
    {2E_{p'+p}(\sqrt{s}-E_{p}-E_{p'}-E_{p'+p})+i\epsilon}
    \quad    \text{for} \quad
    i,j\in {\rm HB}
    \,, 
    \label{btilde}
\end{align}
with the angular structures of P- and S-wave isobars,
\begin{align}
    v_{i}(p,q)
    =
    \left\{
    \begin{matrix}
    \begin{array}{ll}
        g_i\,(p-q)^{\mu}\,\epsilon_{\mu}(\bm{p}+\bm{q}, \lambda)
        \hfill &\quad \text{for $\ell=1$ isobars,}\\
        1&\quad \text{for $\ell=0$ isobars.}
    \end{array}
    \end{matrix}
    \right.
    \label{eq:v}
\end{align}
The incoming $\rho$-meson, which is not in its center of mass, is parameterized through the polarization vector $\epsilon$~\cite{Chung:1971ri} while the outgoing one is expressed through the complex conjugate, $\epsilon^*$. Helicities are indicated with $\lambda\in\{1,0,-1\}$ with each state corresponding to a channel, $\lambda=\lambda(i)$. Explicit expressions can be found, e.g., in the Appendix A in Ref.~\cite{Sadasivan:2021emk}. In this paper, the $\rho$ and $K^*$ mass in $\epsilon$ are chosen as $m_\rho=m_{K^\star}=6.45\,m_\pi$. 
As these particles are resonances, this quantity is only approximately defined and we chose it mass degenerate. Results depend on the choice only very weakly, as tested.
For a connection of the $\tilde B$ term to Lagrangians see App.~\ref{sec:lagrangians}. The coupling constants $g_i$ are discussed in Sec.~\ref{sec:props}.

The $\tilde B$ term contains the angular structure of the two-body sub-amplitudes; in the current convention, the latter have their isospin structure absorbed in the definition of the propagators $\tilde\tau$ defined below. This allows one to express the isospin coefficients $\tilde I_F$ as simple re-coupling of Clebsch-Gordan coefficients - we anticipate in Eq.~\eqref{btilde} the final result given by the matrix elements of Eq.~\eqref{isoexplicit}, $(\tilde {\bm I}_F)_{ji}$. 
To derive that result, we start with the case without strangeness. With the labeling of third components of the isospin according to Fig.~\ref{fig:exlabels}, one finds
\begin{align}
    \tilde I_F&=\sum_{\substack{m,n\\m',n'} }
    \cg{I_In}{I_Sm}{II_3}
    \cg{I_{I'}n'}{I_{S'}m'}{II_3}
    \cg{I_xn-m'}{I_{S}m}{I_{I'}n'} 
    \cg{I_xn-m'}{I_{S'}m'}{I_In}
    \ , \quad \text{ for $\pi\pi\pi$ only.} 
    \label{eq:isofac}
\end{align}
where $I_{I}$ ($I_{I'}$) is the isospin of the incoming (outgoing) isobar,  $I_S$ ($I_{S'}$) is the isospin of the incoming (outgoing) spectator, and $I_x$ is the isospin of the exchanged particle. Equation~\eqref{eq:isofac} can be illustrated as follows. The two-body input of the current formalism are
%} 
isobar amplitudes in the isospin basis. To construct the particle exchange, one needs to first ``unfold'' these amplitudes to the particle/charge basis (third and fourth coefficient). One then has to sum over all possible particle exchanges. Finally, one needs to construct total isospin states (first and second coefficient). The particle phases associated with $\pi^+$ and $\rho^+$ always cancel in the process.

The clear separation of the isospin structure associated with the exchange, cast into $\tilde I_F$, from the isobar propagation is different from previous work~\cite{Sadasivan:2020syi, Mai:2021nul, Sadasivan:2021emk}. There, Lagrangians were used to calculate isospin coefficients leading to different values, because they partly contained isospin structure from the isobar propagation, see the discussion in Appendix \ref{sec:lagrangians}. We, therefore, introduce the ``tilde'' notation here to distinguish it from previous work. For example, the $\pi\rho\to\pi\rho$ transition $\tilde B$ is half the size of the $B$ of these previous studies; accordingly, a factor of two is absorbed in $\tilde \tau$ compared to the previous $\tau$ (for explicit expressions of $\tilde\tau$, see Sec.~\ref{sec:tau} below).

%%%%%%%%%%%%%%%%%%%%%%%

\subsubsection{Inclusion of strangeness}

For strangeness exchange, the situation is slightly more complicated due to definite $G$-parity eigenstates of Eq.~\eqref{gparity}.
Also, we adopt the convention of ``negative strangeness first'' regarding the order of the first two arguments of the CG coefficients of a given vertex, e.g., $\cg{I_xn-m'}{I_{S'}m'}{I_In}$. If a positive-strangeness particle is exchanged, this means for integer $I_I$ a factor of 
\begin{align}
 (-1)^{I_I-I_x-I_{S'}}=(-1)^{I_I-1}=\begin{cases}
     -1 &\text{for\,\,} I_I=0 \ ,\\
     1  \ &\text{for\,\,} I_I=1 \ ,     
 \end{cases}   
\end{align}
and, accordingly, for the case of an outgoing isobar. For $I_I=1$, we have a vector meson with momentum structure of the vertex $p-p'$. Therefore, if one exchanges the order of particles, one gets a minus sign. Overall, if a particle with positive strangeness is exchanged, $S_X=+1$, and if we also have an isobar with zero strangeness ($\sigma$ or $\rho$), then a minus sign should be inserted for that vertex.
Of course, if one uses Lagrangians to calculate the isospin coefficients, this result would arise automatically as we have confirmed. Note that we do not have to consider the case $I_I=2$ because one cannot have $I_X=1/2$ exchange for that case - that transition will vanish, anyway. We, thus, get additional factors of
 \begin{align}
     \tilde c(S_X, S_I)=
     \begin{cases}
         - 1&\text{if } S_X=\mp 1 \text{ and } S_I=0 \ ,\\
         1&\text{else.}
     \end{cases}
     \label{ctilde}
 \end{align}
Here, the upper minus sign corresponds to a ``positive-strangeness-first'' convention and the lower plus sign to a ``negative-strangeness-first'' convention. This only affects the sign of off-diagonal transitions in $B$, and therefore also the full $T$. No observable depends on the choice. If one uses Lagrangians to determine the $\tilde I_f$, that sign is fixed, see Appendix~\ref{sec:lagrangians}.
 
Altogether, for $\eta_G=\pm 1$ we obtain for isobars of strangeness $S_I$ and $S_{I'}$:
\begin{align}
\tilde I_F=\sum_{\substack{m,n\\m',n'} }
\sum_{\substack{S_I, S_{I'},\\ S_X\in\{-1,0,1\}}}
\sum_{I_X\in\{\nicefrac{1}{2},1\}}
&
c(\eta_G,S_I) c(\eta_G,S_{I'})
\delta_{S_I,S_X-S_{I'}}
\cg{I_In}{I_Sm}{II_3}
\tilde c(S_X,S_I)
\cg{I_xn-m'}{I_{S'}m'}{I_In}
 \nonumber \\
\times
&
\tilde c(S_X,S_{I'})
\cg{I_xn-m'}{I_{S}m}{I_{I'}n'}
%\delta_{S_{I'},S_X-S_{I}}
\cg{I_{I'}n'}{I_{S'}m'}{II_3}
\ .
\label{eq:isofacstrange}
\end{align}
The expression contains now also a sum over all values of strangeness and isospin of the exchanged particle. The sums over $S_I$ and $S_{I'}$ in Eq.~\eqref{eq:isofacstrange} extend over the values $-1,1$ for strange isobar and zero for non-strange isobar.
The coefficients for given $G$-parity, reflecting the combination of Eq.~\eqref{gparity}, are 
\begin{align}
c(1,-1)=\frac{1}{\sqrt{2}},\,\,
c(1,1)=\frac{1}{\sqrt{2}},\,\,
c(-1,-1)=\frac{1}{\sqrt{2}},\,\,
c(-1,1)=-\frac{1}{\sqrt{2}},\,\,
c(\pm 1,0) =1 \ .
\label{eq:wstrange}
\end{align}
We explicitly checked that with Eq.~\eqref{eq:wstrange} the transitions for $\eta_G=+1$ to and from the $3\pi$ channels vanish because $\eta_G=-1$ for three pions. 
Furthermore, we checked Eq.~\eqref{eq:isofacstrange} by calculating exchange diagrams with Lagrangians as explained in Appendix~\ref{sec:lagrangians}.

The resulting coefficients for different $G$-parities, isospins, and coupled channels are shown in Tables~\ref{tab:I=0 IFtilde}, \ref{tab:tildeiso1}, and \ref{tab:I=2 IFtilde}. The abbreviation $ K_{\nicefrac{1}{2}}$ stands for the $I_I=\nicefrac{1}{2}$ isobars $\kappa$ or $K^*(892)$ and $K_{\nicefrac{3}{2}}$ for the $I_I=3/2$ S-wave $\pi K$ isobar. The lower (upper) sign corresponds to the ``Positive (negative)-strangeness first'' convention for the exchanged particle as discussed before. 

The coefficients only depend on isospin, but in a given coupled-channel scheme some transitions are still forbidden: for example, with the channel ordering of Table~\ref{tab:channels}, there are no $\pi\rho\to KK^*$ transitions as the $K\bar K$ channel for $\rho$ quantum numbers is neglected, as discussed in Sec.~\ref{sec:channels}. In this approach, we allocate these forbidden transitions in the definitions of the isospin factors. In the JLS basis, we can then write the isospin $I=1$, $\eta_G=-1$ factors in symmetric matrix form as
\begin{align}
   \tilde {\bm I}_f (I=1,\eta_G=-1)=
   \ \ 
   \begin{blockarray}{ccccccccc}
   (\pi\rho)_S&(\pi\rho)_D&(KK^*)_S&(KK^*)_D&(\pi\sigma)_P&(\pi(K\bar K)_S)_P&(\pi\pi_2)_P&(K\kappa)_P&(KK_{\nicefrac{3}{2}})_P\\
       \begin{block}{(ccccccccc)}
        \nicefrac{1}{2}&\nicefrac{1}{2}&0&0&\nicefrac{1}{\sqrt{3}}&0&-\sqrt{\nicefrac{5}{12}}&0&0\\
        & \nicefrac{1}{2}&0&0&\nicefrac{1}{\sqrt{3}}&0&-\sqrt{\nicefrac{5}{12}}
        & 0 & 0         \\
        && -\nicefrac{1}{3} &-\nicefrac{1}{3} &0& -\sqrt{\nicefrac{2}{3}} & 0
        &-\nicefrac{1}{3} &-\sqrt{\nicefrac{8}{3}}
        \\
        &&& -\nicefrac{1}{3} &0& -\sqrt{\nicefrac{2}{3}} & 0
        &-\nicefrac{1}{3} &-\sqrt{\nicefrac{8}{3}}
        \\
        &&&& \nicefrac{1}{3} & 0 & \nicefrac{\sqrt{5}}{3} &0&0        \\
        &&&& & 0 & 0 &-\sqrt{\nicefrac{2}{3}}
        &-\nicefrac{2}{\sqrt{3}}
        \\   
        &&&&&& \nicefrac{1}{6} &0 & 0
        \\
        &&&&&&& -\nicefrac{1}{3}&-\sqrt{\nicefrac{8}{3}}
        \\
        &&&&&&&& \nicefrac{1}{3}
        \\
    \end{block}
    \end{blockarray}\ \ ,
    \label{isoexplicit}
\end{align}
where we decided to use the "positive-strangeness first" convention (lower sign in Table~\ref{tab:tildeiso1}), because these results coincide with the Lagrangian method of App.~\ref{sec:lagrangians}.

%%%%%%%%%%%%%%%%%%%%%%%%%%%%%%%%%%%%%%%%%%%%%%

\begin{table}[htb]
\center
    \begin{tabular}{c|cc}
    \hline\hline
    $\eta_G=-1, I=0$
   & $\pi\rho$    & 
    $K K_{\nicefrac{1}{2}}$  \TT\BBB \\
    \hline
$\pi\rho$   & $-1$ & $\mp\sqrt{2}$ 
\TT \\
$K K_{\nicefrac{1}{2}}$   & & $-1$\BBB\\
\hline
\hline
    \end{tabular}
    %%%%%%%%%%%%%%%%%%%%%%%%%%%%%%%%%%%%%%%%%%%%    
     \hspace*{1cm}
    \begin{tabular}{c|cc}
    \hline\hline
    $\eta_G=+1, I=0$
   & 
    $K K_{\nicefrac{1}{2}}$  \TT\BBB \\
    \hline
\TT 
$K K_{\nicefrac{1}{2}}$   &  $1$\BBB\\
\hline
\hline
    \end{tabular}
    \caption{  Isospin factors $\tilde I_F$ for total isospin $I=0$ and different $G$-Parities.
    The abbreviation $\pi_2$ means the repulsive $(I_I,\ell)=(2,0)$ channel, $K_{\nicefrac{1}{2}}$ stands for the isospin $I_I=\nicefrac{1}{2}$ isobars $\kappa$ or $K^*(892)$. The lower (upper) sign corresponds to a "positive  (negative) strangeness first" convention in the calculation of the $\tilde I_F$, see text.}
    \label{tab:I=0 IFtilde}
\end{table}

\begin{table}[htb]
\center
    \begin{tabular}{c|ccccc}
    \hline\hline
$\eta_G=-1, I=1$   & $\pi\rho$   & $\pi f_0$ & $\pi\pi_2$ &
    $K K_{\nicefrac{1}{2}}$ & $K K_{\nicefrac{3}{2}}$ \TT \BBB\\
    \hline
$\pi\rho$   & $\nicefrac{1}{2}$&$\nicefrac{1}{\sqrt{3}}$&$-\sqrt{\nicefrac{5}{12}}$&$\pm\nicefrac{2}{\sqrt{3}}$&$\mp\sqrt{\nicefrac{2}{3}}$
\TT \\
$\pi f_0$   & &$\nicefrac{1}{3}$&$\nicefrac{\sqrt{5}}{3}$&$\pm\sqrt{\nicefrac{2}{3}}$& $\pm\nicefrac{2}{\sqrt{3}}$ \\
$\pi\pi_2$   & &&$\nicefrac{1}{6}$&$0$&$0$\\
$K K_{\nicefrac{1}{2}}$  & &&&$-\nicefrac{1}{3}$&$-\sqrt{\nicefrac{8}{3}}$\\
$K K_{\nicefrac{3}{2}}$  & &&&&$\nicefrac{1}{3}$ \BBB \\
\hline
\hline
    \end{tabular}    
%%%%%%%%%%%%%%%%%%%%%%%%%%%%%%%%%%%%%%%%%%%%    
     \hspace*{1cm}
        \begin{tabular}{c|cc}
    \hline\hline
$\eta_G=+1, I=1$    &
    $K K_{\nicefrac{1}{2}}$ & $K K_{\nicefrac{3}{2}}$ \TT \BBB \\
    \hline
$K K_{\nicefrac{1}{2}}$  & $\nicefrac{1}{3}$&$\sqrt{\nicefrac{8}{3}}$\\
$K K_{\nicefrac{3}{2}}$  & & $-\nicefrac{1}{3}$ \BBB \\
\hline
\hline
    \end{tabular}
    \caption{Isospin factors $\tilde I_F$ for total isospin $I=1$ and different $G$-Parities. The notation $K_{\nicefrac{3}{2}}$ stands for the repulsive $\pi K$ isobar with $(I_I,\ell)=(\nicefrac{3}{2},0)$. 
    The symbol $f_0$ stands two pions or $K\bar K$ in $(I_I,\ell)=(0,0)$, and $\pi_2$ for two pions in $(I_I,\ell)=(2,0)$.
    See Table~\ref{tab:I=0 IFtilde} for other abbreviations.  Note that not all transitions are always realized owing to the coupled-channel structure of the isobars themselves, see Eq.~\eqref{isoexplicit}.}
    \label{tab:tildeiso1}
\end{table}

%%%%%%%%%%%%%%

\begin{table}[htb]
\center
    \begin{tabular}{c|ccc}
    \hline\hline
$\eta_G=-1, I=2$   & $\pi\rho$    & $\pi\pi_2$ &
    $K K_{\nicefrac{3}{2}}$ \TT \BBB \\ \hline
$\pi\rho$   & $\nicefrac{1}{2}$&$\nicefrac{\sqrt{3}}{2}$&$\pm\sqrt{2}$
\TT \\
$\pi\pi_2$   & &$-\nicefrac{1}{2}$& $0$\\
$K K_{\nicefrac{3}{2}}$   & && $-1$ \BBB \\
\hline
\hline
    \end{tabular}
    %%%%%%%%%%%%%%%%%%%%%%%%%%%%%%%%%%%%%%%%%%%%    
     \hspace*{0.5cm}
    \begin{tabular}{c|cc}
    \hline\hline
    $\eta_G=+1, I=2$
   & 
    $K K_{\nicefrac{3}{2}}$  \TT\BBB \\
    \hline
\TT 
$K K_{\nicefrac{3}{2}}$   &  1\BBB\\
\hline
\hline
    \end{tabular} 
        %%%%%%%%%%%%%%%%%%%%%%%%%%%%%%%%%%%%%%%%%%%%    
     \hspace*{0.5cm}
    \begin{tabular}{c|cc}
    \hline\hline
    $\eta_G=-1, I=3$
   & 
    $\pi\pi_2$  \TT\BBB \\
    \hline
\TT 
$\pi\pi_2$   &  $1$\BBB\\
\hline
\hline
    \end{tabular}   
    \caption{Isospin factors $\tilde I_F$ for total isospins $I=2$, $I=3$, and different $G$-Parities. See Tables~\ref{tab:I=0 IFtilde} and \ref{tab:tildeiso1} for abbreviations.}
    \label{tab:I=2 IFtilde}
\end{table}

%%%%%%%%%%%%%%%%%%%%%%%%%%%%%%
\subsection{Isobar-spectator propagation and symmetry factors}
\label{sec:props}
The scattering equation in the JLS basis is formulated in the product space of nine channels and complex momenta. With the channel ordering of Table~\ref{tab:channels}, the isobar propagator $\tilde\tau$ of Eq.~\eqref{eq:T3-integral-equation} acquires a block-diagonal structure, 
\label{sec:tau}
\begin{align}
    \tilde\tau=\text{diag}\left(
        \tilde\tau_{\pi\rho},
        \tilde \tau_{\pi\rho},
        \tilde\tau_{KK^*},
        \tilde\tau_{KK^*},
        \begin{pmatrix}
        \tilde \tau^0_{\pi\pi} & \tilde \tau^0_{\pi K} \\
        \tilde \tau^0_{K\pi} & \tilde \tau^0_{K\bar K}
        \end{pmatrix},
        \tilde \tau_{\pi\pi_2},
        \tilde \tau_{K\kappa},
        \tilde \tau_{KK_{\nicefrac{3}{2}}}
    \right) \ ,
\end{align}
in which each element is an $n\times n$ matrix of complex spectator momenta (see Sec.~\ref{sec:jls} for a discussion of the momentum discretization to solve the scattering equation). Here, $\tilde \tau_{\pi\pi}^0$ stands for the $I_I=0,\,
\ell=0$ $\pi\pi\to\pi\pi$ transition, while the other $\tilde \tau^0$ stand for the corresponding $\pi\pi\to K\bar K$ and $K\bar K\to K\bar K$ transitions.
An example for these coupled-channel transitions is shown in Fig.~\ref{fig:ccillu}. The off-diagonal elements encode the possibility for the channel to change while the isobar is ``in-flight'', due to its coupled-channel nature, see Eq.~\eqref{matchuchpt}.
The matrix $\tilde \tau$ is not affected by the partial-wave projection discussed in the next section.

Regarding the parameterization of individual isobars, one has to adapt different conventions used in the literature, and be aware of how symmetry factors are distributed over different ingredients of the two-body amplitude. This is critical in the two-channel $\pi\pi/K\bar K$ isobar case, because, in the isospin basis, pions count as indistinguishable while $K\bar K$ do not. 

\begin{figure}[htb]
   \begin{minipage}{0.48\textwidth}
     \centering
     \includegraphics[width=1\linewidth]{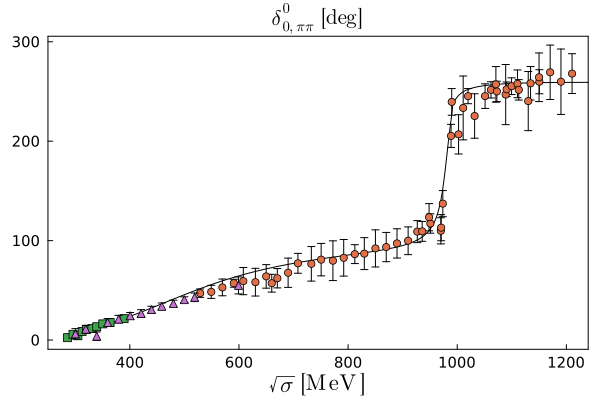}
   \end{minipage}
   \begin{minipage}{0.48\textwidth}
     \centering
     \includegraphics[width=1\linewidth]{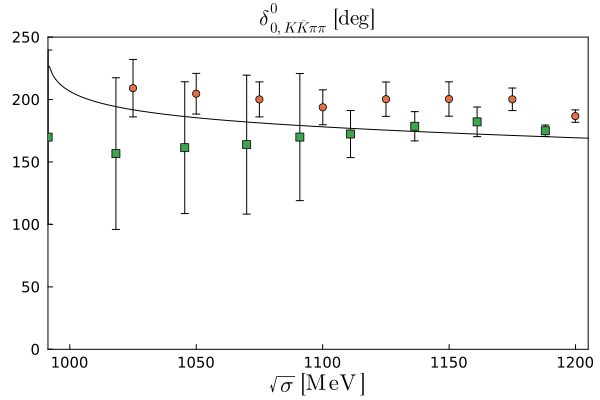}
   \end{minipage} 
   \begin{minipage}{0.48\textwidth}
     \centering
     \includegraphics[width=1\linewidth]{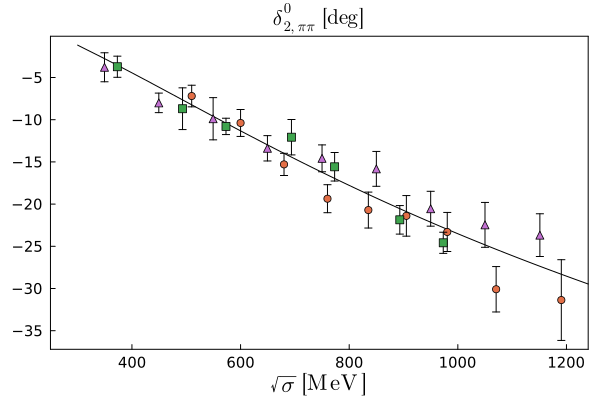}
   \end{minipage} 
   \begin{minipage}{0.48\textwidth}
     \centering
     \includegraphics[width=1\linewidth]{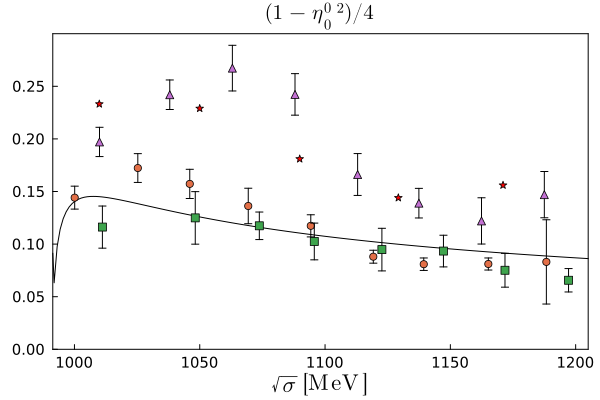}
   \end{minipage} 
   \begin{minipage}{0.48\textwidth}
     \centering
     \includegraphics[width=1\linewidth]{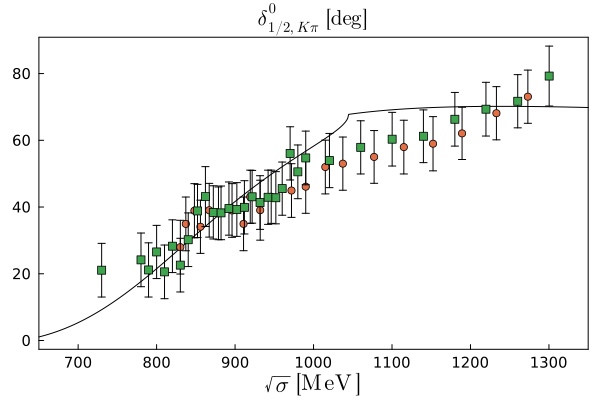}
   \end{minipage}
   \begin{minipage}{0.48\textwidth}
     \centering
     \includegraphics[width=1\linewidth]{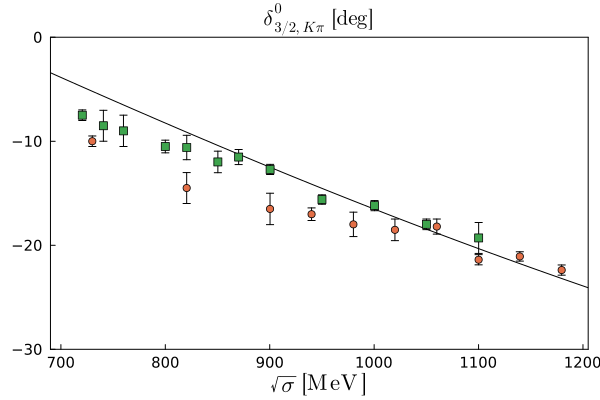}
   \end{minipage} 
   \caption{Parameterizations (solid lines) of the phase shifts  and inelasticity in the notation $\delta_{I_I}^\ell$ and $\eta_{I_I}^\ell$. Data: For $\delta^0_{0,\pi\pi}$: green squares \cite{NA482:2007xvj}, purple triangles \cite{Froggatt:1977hu} and orange circles are averages as defined in \cite{Oller:1998zr}. For $\delta^0_{0,K\bar{K}\pi\pi}$: green squares \cite{Martin:1979gm} and orange circles \cite{Cohen:1980cq}. For $(1-\eta^{02}_{0})/4$: red stars \cite{Froggatt:1977hu}, orange circles \cite{Cohen:1980cq}, green squares \cite{Etkin:1981sg} and purple triangles \cite{Lindenbaum:1991tq}. For $\delta^0_{2,\pi\pi}$: purple triangles \cite{Schenk:1991xe}, green squares \cite{Janssen:1994wn} and orange circles \cite{Rosselet:1976pu}. For $\delta^0_{1/2,K\pi}$: green squares are averages as defined in \cite{Oller:1998zr} and orange circles \cite{Aston:1987ir}. For $\delta^0_{3/2,K\pi}$: green squares \cite{Linglin:1973ci} and orange circles \cite{Estabrooks:1977xe}.
   These parameterizations are used for the scalar two-body input as can be seen in Eq.~\eqref{matchuchpt}.}
   \label{phase_shifts_noV}
\end{figure}

The rule of thumb is that $\tilde \tau$, together with adjacent angle-dependent decay vertices absorbed in the definition of $\tilde B$, should correspond to the full, plane-wave, isospin-projected $2\to 2$ transition amplitude which no longer has symmetry factors absorbed in it:

For the $I_I=0,\ell=0$ isobar we use a convenient parameterization including up to NLO CHPT contact terms~\cite{Oller:1998hw}. While a full NLO transition in SU(3), including $t$- and $u$-channel loops has been formulated~\cite{GomezNicola:2001as}, here we only need an efficient parameterization of the two-channel amplitude. For the sake of a better data description, we even allow for different subtraction constants for different channel loops. We show in Table \ref{tab:subtraction} the different subtraction constants $\alpha^{DR}$ obtained from the fit. Figure~\ref{phase_shifts_noV} shows the comparison of data and fit, including both phases and the inelasticity for the scalar-isoscalar channel. 
\begin{table}[htb]
\center
    \begin{tabular}{c|c|c|c|c|c|c}
    \hline\hline
 $(I_I,\ell)$ & $(0,0)$ & $(1/2,0)$ & $(1/2,1)$ & $(1,1)$ & $(3/2,0)$ & $(2,0)$  \TT \BBB \\ \hline
 Channels & $\pi\pi,\, K\bar K$ & $\pi K,\,\eta K$ &$\pi K,\,\eta K$ & $\pi\pi,\, K\bar K$ & $\pi K$ & $\pi\pi$\\
$\alpha^{DR}$ & $-1.13(9)$ & $-0.00(1)$ & $-0.00(19)$ & $-2.67(13)$ & $-1.58(19)$ & $-2.31(27)$ 
\BBB \\
\hline
\hline
    \end{tabular}
    \caption{Channels and subtraction constants obtained in the fit to two-body input using the model of Ref.~\cite{Oller:1998hw}. For vector $\ell=1$ isobars we use a different (one-channel) parameterization than shown here.}
    \label{tab:subtraction}
\end{table}
%
%%%%%%%%%%%%%%%%%%%%%%%%%%%%%%%%%%%%%%%%%%%%%%%

%%%%%%%%%%%%%%%%%%%%%%%%%%%

The connection of the chiral unitary amplitudes to the isobar-spectator propagators is given as
\begin{align}
\tilde\tau^0_{\pi\pi}=2 T^{00}_{\pi\pi\to \pi\pi} \ ,\quad
\tilde\tau^0_{\pi K}=\tilde\tau^0_{K\pi}=\sqrt{2} T^{00}_{\pi\pi\to K\bar K}\ ,\quad \tilde\tau^0_{K\bar K}= T^{00}_{K \bar K\to K\bar K} \ ,\quad
\tilde\tau_{K\kappa}=T^{{\nicefrac{1}{2}}0}_{\pi K\to \pi K}\ , \quad
\tilde\tau_{K K_{\nicefrac{3}{2}}}=T^{{\nicefrac{3}{2}}0}_{\pi K\to \pi K}
\label{matchuchpt}
\end{align} 
where $T^{I_I\ell}$ are the partial-wave amplitudes of Ref.~\cite{Oller:1998hw}. In particular, in that normalization the symmetry factors are absorbed in the definition of the states (inserted when defining isospin combinations~\cite{Oller:1998hw}); for a full amplitude, these factors have to be removed according to Eq.~\eqref{matchuchpt}. 

For completeness we mention that in the chiral unitary normalization as used for the Inverse Amplitude Method (IAM), symmetry factors are inserted in the transition from plane-wave to partial-wave amplitude  for channel ${a\to b}$~\cite{GomezNicola:2001as}:
\begin{align}
T^{I\ell}_{ab\text{, IAM}}(p',p)=\frac{1}{32N\pi}\int\limits_{-1}^1\diff \cos\theta\,P_\ell(\cos\theta)T^I_{ab\text{, IAM}}(\bm{p}',\bm{p}) \ .
\label{IAMPWA}
 \end{align}
For identical particles in $a$ and $b$, like $\pi\pi\to\pi\pi$ for any isospin, $N=2$; for $\pi\pi\to K\bar K$, $N=\sqrt{2}$; and for $K\bar K\to K\bar K$, $N=1$. 
The inversion is given by 
\be
T^I_{ab\text{, IAM}}(\bm{p}',\bm{p})=64 N\pi^2\sum_{\ell,m}\,
Y_{\ell m}^*(\hat{\bm{p}}')\,T^{I\ell}_{ab\text{, IAM}}(p',p)\,Y_{\ell m}(\hat{\bm{p}}) \ ,
\label{TIAMplane}
\ee
which makes the connection to the isobar-spectator propagator apparent,
\begin{align}
\tilde \tau_{\pi\sigma}&=-32\pi T_{\pi\pi\to \pi\pi \text{, IAM}}^{00}    \ ,
\label{eq:moretildetaus}
\end{align}
for the example of isoscalar $\pi\pi\to\pi\pi$, 
taking into account an additional difference in sign for the $T$-matrix in IAM normalization, namely that (in operator notation) $S=1+iT$ for IAM while $S=1-iT$  in the present framework and also for the two-body input from Ref.~\cite{Oller:1998hw}. Note that, in any case, the imaginary parts of the $\tilde \tau^{-1}$ are in agreement with three-body unitarity, whether they come from IAM, Eq.~\eqref{matchuchpt} or the isobar self energy discussed in the following.

For the P-wave isobars, one could employ the same chiral unitary framework as for the other channels~\cite{Oller:1998hw}, after transforming them to the plane-wave basis. 
Here, we simplify the two-channel formulation of these vector channels (see Table~\ref{tab:subtraction}) to one channel due to the small inelasticities in these channels. Using s-channel propagators for the interaction leads to~\cite{Sadasivan:2021emk}
\begin{align}        \tilde\tau_{V,ji}=\frac{I(VM_jM_j')I(VM_iM_i')}{\sigma-m_V^2-\Sigma_V}
\label{eq:tauv}
\end{align}
where $V$ ($M$) stands for vector (pseudoscalar) mesons,  $I(VMM')$ is the isospin factor for the $V\to MM'$ transition calculated in App.~\ref{sec:lagrangians}, $I(\rho\pi\pi)=\sqrt{2}$, $I(K^*\pi K)=\sqrt{\nicefrac{3}{4}}$, and $I(\rho K\bar K)=1/\sqrt{2}$ which is not used as we neglect that decay, i.e., the above $\tilde\tau_V$ is effectively one-channel. Above, $m_V$ is a fit parameter, and $\Sigma_V$ is the self energy, generalized to the unequal mass case. 
\begin{align}
\Sigma_{V}=\int_0^\infty\frac{dl\,l^2}{(2\pi)^3}\frac{E_1+E_2}{2NE_1 E_2}
\left(\frac{\sigma}{\sigma'}\right)^2
\frac{\tilde v_V^2(l)}{\sigma-\sigma^{\prime 2}+i\epsilon} \ ,
\quad\text{where }\,
\sigma'=(E_1+E_2)^2 \ ,
\label{eq:selfun}
\end{align}
and $E_i^2=m_i^2+l^2$ are the energies of the decay products of the isobar, $N=2$ for identical decay products (i.e., $\pi\pi$), and $N=1$ for non-identical particles (i.e., $\pi K$ and $K\bar K$). In the identical-particle case, this reduces to the self energy of Ref.~\cite{Sadasivan:2021emk}. In Eq.~\eqref{eq:selfun}, the P-wave projected vertex is given by
\begin{align}
    \tilde v_{VMM'}(q)=I(VMM')\,\sqrt{\frac{16\pi}{3}}\,g_V\,q 
    \label{eq:vpw}
    \ .
\end{align}
The coupling constant $g_V$ is another fit parameter, and extensions to more subtractions along the lines of Ref.~\cite{Sadasivan:2021emk} are possible. Note that $g_V=g_i$ from Eq.~\eqref{eq:v} are the same constants.

\begin{figure}[htb]
    \begin{minipage}{0.48\textwidth}
     \centering
     \includegraphics[width=1\linewidth]{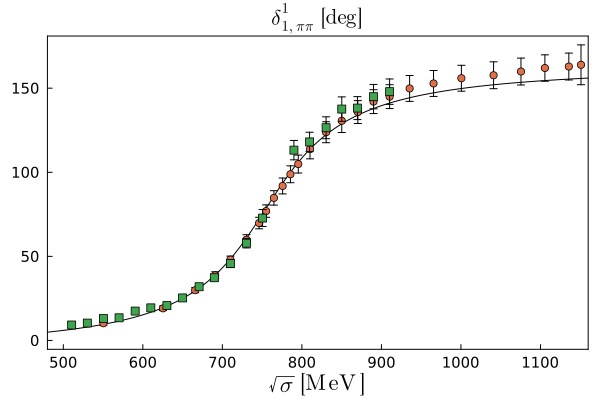}
   \end{minipage} 
   \begin{minipage}{0.48\textwidth}
     \centering
     \includegraphics[width=1\linewidth]{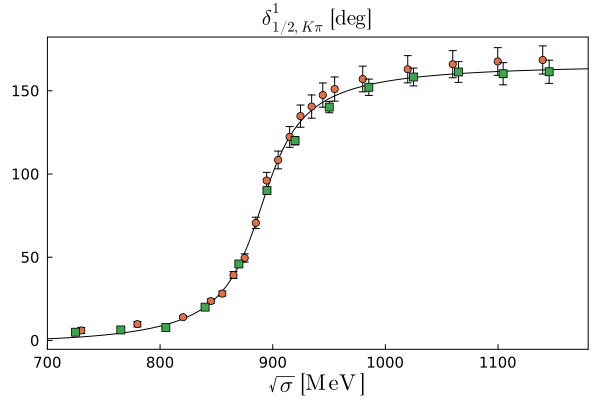}
   \end{minipage} 
   \caption{Parameterizations (solid lines) of the phase shifts for vector mesons. Data: For $\delta^1_{1,\pi\pi}$: orange circles \cite{Estabrooks:1974vu} and green squares \cite{Lindenbaum:1991tq}. For $\delta^1_{1/2, K\pi}$: orange circles \cite{Estabrooks:1977xe} and green squares \cite{Mercer:1971kn}.
   These parameterizations are used for the vector two-body input $\tilde{\tau}_V$ as defined  in Eq.~\eqref{eq:tauv}.}
   
\label{phase_shifts_V}
\end{figure}
 With the partial-wave amplitude
 \begin{align}
T^{I_I1}(q,p)=\frac{\tilde v_V(q)\tilde v_V(p)}{\sigma-m_V^2-\Sigma_V}
 \end{align}
 in isospin $I_I=1$ and $I_I=1/2$, 
the  $\delta_{I_I}^{1}$ phase shifts can be fitted with the result shown in Fig.~\ref{phase_shifts_V} with 
$g_\rho=5.89$, $m_\rho=5.51~m_\pi$,
$g_{K^*}=7.08$ and $m_{K^*}=6.72~m_\pi$. We do not attempt to propagate any uncertainties from the two-body input to three-body amplitudes, so we consider only central values of the fits.

%%%%%%%%%%%%%%%%%%%%%%%%%%%%%%

\subsection{Partial-wave decomposition and the JLS basis}
\label{sec:jls}
The partial-wave decomposition of the scattering equation~\eqref{eq:T3-integral-equation} proceeds in the same way as in Refs.~\cite{Sadasivan:2020syi, Mai:2021nul, Sadasivan:2021emk}. In particular, the partial-wave projection is obtained by 
\begin{align}
    \tilde B_{ji}(s,p',p)=2\pi\int_{-1}^1 dz\, d^J_{\lambda(j)\lambda(i)}(z)\tilde B_{ji}(s,\bm{p}',\bm{p})\,,
    \label{eq:b pw projection}
\end{align}
where $z=\cos\theta$ for $\theta$ being the scattering angle between $\bm{p}$ and $\bm{p}'$ and $\lambda(i)$ is the isobar helicity of channel $i$. Note that for the spinless $\pi \pi_2, K\kappa,\dots$ channels, $\lambda=0$. Note that in the present convention the isobar is incoming in the $-z$-direction while the outgoing isobar is on the $xz$-plane with negative $x$-component. The standard expression for the partial-wave decomposition of particles with spin is 
\begin{align}
\langle{\lambda_3,\lambda_4, s, \bm{p}'|\tilde B|\lambda_1,\lambda_2, s, \bm{p}\rangle}
    =
    \frac{1}{4\pi}
    \sum_J (2J+1)
d_{\lambda\lambda'}^J(\theta)^*\,
\langle{\lambda_3,\lambda_4, s, p'|\tilde B|\lambda_1,\lambda_2, s, p\rangle} \ ,
\label{genpwa}
\end{align}
where we have chosen $\phi=0$, i.e., $D_{\lambda\lambda'}^J(0,\theta,0)=d_{\lambda\lambda'}^J(\theta)$. 
The quantity $\lambda_3$ ($\lambda_1$) is the helicity of the particle with momentum $\bm{p}'$  ($\bm{p}$), and $\lambda:=\lambda_1-\lambda_2$, $\lambda':=\lambda_3-\lambda_4$. In particular, the first lower index of the Wigner $d$-function in Eq.~\eqref{genpwa} is the incoming helicity difference. As we chose the incoming $\rho$ in the $-z$-direction, its helicity corresponds to $\lambda_2$, and similarly for the outgoing state, it corresponds to $\lambda_4$ (we always have spinless spectators, $\lambda_1=\lambda_3=0$). Using that $d^J_{\lambda'\lambda}=d^J_{-\lambda-\lambda'}$ it becomes apparent that the first index in Eq.~\eqref{eq:b pw projection} corresponds to the outgoing helicity, in our particular convention.

In the JLS basis the $\pi\rho$ and $KK^*$ channels couple to $J=1$ in relative S- and D-wave, while the spinless ($\pi\pi_2,\dots$) channels couple in P-wave according to Table~\ref{tab:channels}. The transformation between HB and JLS bases follows from Ref~\cite{Chung:1971ri}, 
\begin{align}
    \tilde B_{L'L}(s,p',p)&=U_{L'\lambda(j)}\tilde B_{ji}(s,p',p)U_{\lambda(i)L} \ ,
    \nonumber
    \\
    U_{L\lambda }
    =\sqrt{\frac{2L+1}{2J+1}}&\sum_{\lambda_2}\cg{L0}{S\lambda}{J\lambda}\cg{S_1\lambda_1}{S_2-\lambda_2}{S\lambda} 
    = \sqrt{\frac{2L+1}{2J+1}}\cg{L0}{S\lambda}{J\lambda} \ ,
    \label{umat}
\end{align}
where $S_1=S$ ($\lambda_1=\lambda$) is the spin (helicity) of the isobar and $S_2=0$ ($\lambda_2=0$) is the spin (helicity) of the spectator. Note that for $\pi\rho$ and $KK^*$,
\begin{align}
    U=
    \begin{pmatrix}
        \frac{1}{\sqrt{3}}  & \frac{1}{\sqrt{3}}  &\frac{1}{\sqrt{3}}  \\
         \frac{1}{\sqrt{6}}  &-\sqrt{\frac{2}{3}}   &\frac{1}{\sqrt{6}} 
    \end{pmatrix} \,,
    \label{Umatrix}
\end{align}
where the first and second rows denote S- and  D-wave, respectively. 
For the channels without spin,  $U=1$.

After projection of Eq.~\eqref{eq:T3-integral-equation} the isobar-spectator equation reads
\begin{align}
    \tilde T_{ji}(s,p',p)&
    =\tilde B_{ji}(s,p',p)+\tilde C_{ji}(s,p',p)+
    \int\limits_0^\Lambda 
    \frac{\text{d}l\,l^2}{(2\pi)^3\,2E_l}
    \left(\tilde B_{jk}(s,p',l)+\tilde C_{jk}(s,p',l)\right)\,
    \tilde \tau_{k}(\sigma_l) \,
    \tilde T_{kj}(s,l,p) \ ,
    \label{eq:TLL}
\end{align}
where $i,j,k$ are the channel indices in the JLS basis according to Table~\ref{tab:channels}.

%%%%%%%%%%%%%%%%%%%%%%%%%

\subsection{Solution of the integral equation}
\label{sec:ccmethod}
\begin{figure}[htb]
\begin{center}
\includegraphics[width=0.45\textwidth]{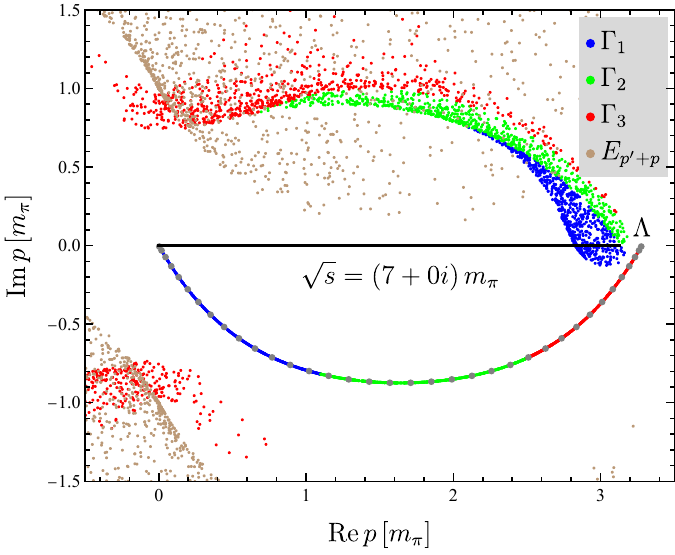}
\end{center}
\caption{An example for a complex contour $\Gamma_{\text{SMC}}$ (blue-green-red line). The blue, green, and red dots indicate $p_{\pm}$ from  Eq.~\eqref{eq: singularity p'} taking this SMC as input for random $x\in[-1,1]$. 
The brown dots show the corresponding results for the pre-factor $E_{p'+p}$ appearing in $\tilde B$ according to Eq.~\eqref{btilde}, i.e., $p(p')$ for $E_{p'+p}=0$ and $p'\in\Gamma_{SMC}$.
The black line indicates physical momenta.}
\label{fig:SMC}
\end{figure}

The integral equation~\eqref{eq:TLL} can be solved along a complex spectator momentum contour (SMC)~\cite{Hetherington:1965zza} as explained in the following, or by direct inversion as explained in Sec.~\ref{sec:direct}.
To summarize the complex SMC method we follow Refs.~\cite{Sadasivan:2020syi, Sadasivan:2021emk}. The integration from $0$ to $\Lambda$ in Eq.~\eqref{eq:TLL} has to fully cover the physical region indicated with the black solid line in Fig.~\ref{fig:SMC}, that is, $\Lambda>q'_\text{max}$ from Eq.~\eqref{eq:qpmx}, for all channels of the problem.
The integration over the physical range conflicts with three-body singularities given by the zeros of the denominator of Eq.~\eqref{btilde}. One can, therefore, deform the momentum integration to the complex contour shown in Fig.~\ref{fig:SMC}. As explained in the caption, the dots indicate solutions of vanishing denominator of $\tilde B$. An overlap of dotted region and SMC would, therefore, indicate a combination of $p,p'$ for which $\tilde B$ becomes singular. Clearly, the complex SMC avoids this scenario. The figure, however, also shows an overlap of the dotted region and parts of the physical region (black solid line). Therefore, if one wishes to obtain the scattering amplitude $\tilde T$ for real spectator momenta, one has to take a few additional steps as explained in the following subsections. For additional details regarding the complex SMC method, see Ref.~\cite{Sadasivan:2021emk}.

To solve Eq.~\eqref{eq:TLL}, the integration momentum is discretized along the SMC. This allows one to convert the integral equation into a matrix equation that can be easily solved for  $\tilde T$ at these discrete incoming and outgoing momenta~\cite{Sadasivan:2020syi}. 
In this work, we use the complex SMC 
\begin{align}
    \Gamma_{\text{SMC}}(t)=t+i\, V_0 (1-e^{-t/w})(1-e^{(t-\Lambda)/w})\,,
    \label{eq: SMC function}
\end{align}
where the parameters are set to be $V_0=-1, \,w=0.6$ and $t\in[0,\Lambda]$.  In the numerical calculation, we chose a  total energy of $\sqrt{s}=7\,m_\pi$ and $\Lambda=3.28\,m_\pi$ unless indicated otherwise.

If one searches for resonance poles, one has to additionally allow for complex $\sqrt{s}$  with the appropriate analytic continuation. A detailed discussion on the analytic structure can be found in Ref.~\cite{Sadasivan:2021emk}. In particular, with the right choice of the SMC reaching sufficiently into the complex plane, one can directly dial in complex energy values in a large range to search for poles with little or no additional effort~\cite{Sadasivan:2021emk}.

While resonance poles appear at the same position irrespective of whether the momentum is real or complex, production reactions require final-state interactions in form of a ``half-real'' $T$-matrix. This situation is illustrated in Fig.~\ref{fig:ccillu}. 
For that specific example, we need to know $T$ for real-valued outgoing spectator momenta $q'$ of a $\pi\rho$ channel, and for complex-valued, incoming spectator momenta $p$ of a $KK^*$ channel; complex-valued, because there is an integration over the initial $K^*K\bar K$ triangle, for which one can re-use the same complex contour employed in the solution of the integral equation. The calculation of half-real $T$-matrices and production amplitudes is described in Sec.~\ref{sec:anal}.

%%%%%%%%%%%%%%%%%%%%%%%%%%%%%%%%%%%%%%%%%%%%%%%%%%%%%%%

\subsection{Amplitude for real momenta through contour deformation}
\label{sec:anal}
%\com{MD}{Copied here from the pheno lab diary; abandon there, only keep this version updated.}
\begin{figure}[htb]
\begin{center}
\includegraphics[width=0.55\textwidth]{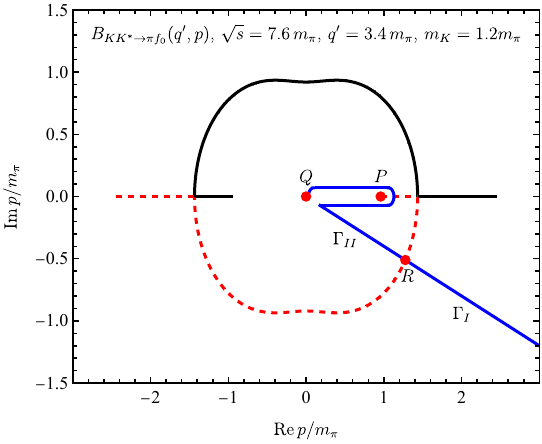}
\end{center}
\caption{Illustration of the integral path of Eq.~\eqref{eq:fancygamma} in the "critical region" for fixed outgoing $q'$, shown in the complex plane of incoming momentum $p$. The black and red dashed lines indicate the two cuts according to Eq.~\eqref{eq: singularity3B}. The integration starts at the origin on the first Riemann sheet. Then, it goes around the logarithmic singularity at $P$, passing through the cut down onto the second Riemann sheet, and back to  $Q$, that indicates the origin on the second Riemann sheet. From there, the integration follows the original SMC $\Gamma$ with a smooth analytic transition of $\tilde B$ at $R$ from second sheet, $B^{II}$, back to the first sheet, $B^I$.}
\label{fig:illustration}
\end{figure}

The integral equation~\eqref{eq:TLL} is solved along a complex contour as discussed.
At a fixed three-body energy $\sqrt{s}$ the position in incoming momentum $p$, where the denominator of $\tilde B_{ji}(s,\bm{q}',\bm{p})$ (from Eq.~\eqref{btilde}), % given by
\begin{align}
f(\bm{q}',\bm{p})=\sqrt{s}-E_{q'}-E_p-E_{q'+p}    \ ,
\label{eq:ffunc}
\end{align}
vanishes is given by 
\begin{align}
&p_{\pm}(m_1,m_3,m_u,x)\nonumber \\
&=\frac{q'x(q'^2-\alpha^2+m_u^2-m_1^2)\pm\alpha\sqrt{\left(\beta-m_1\right)^2-m_u^2+q^{\prime 2}\left(x^2-1\right)}\sqrt{\left(\beta+m_1\right)^2-m_u^2+q^{\prime 2}\left(x^2-1\right)}}{2\beta^2}
\label{eq: singularity3B}
\end{align}
where
\begin{align}
\alpha=\sqrt{s}-\sqrt{m_3^2+q^{\prime 2}}\ ,\quad\beta^2=\alpha^2-q^{\prime 2}x^2 \ ,\nonumber
\end{align}
for incoming (outgoing) spectator of mass $m_1$ ($m_3$) and exchanged mass $m_u$. This expression
simplifies to
\begin{align}
p_{\pm} (m_\pi,m_\pi,m_\pi)=\frac{q'x(q'^2-\alpha^2)\pm\alpha\sqrt{\left(\beta^2+q^{\prime 2}(x^2-1)\right)^2-4m^2_\pi\beta^2}}{2\beta^2}
\label{eq: singularity p'}
\end{align}
for the case of three pions.
As indicated in Fig.~\ref{fig:ccillu}, we denote the outgoing momentum as $q'$ when it is real, and as $p'$ when it is complex. The incoming momentum $p$ is supposed to be always complex. There is also a factor of  $E_{q'+p}$ in the denominator of Eq.~\eqref{btilde} that can vanish for suitable combinations of complex momenta, but this never induces problems.

\subsubsection{Location of the critical region}
\label{sec:critical}
\begin{figure}[htb]
\begin{center}
\includegraphics[width=0.55\textwidth]{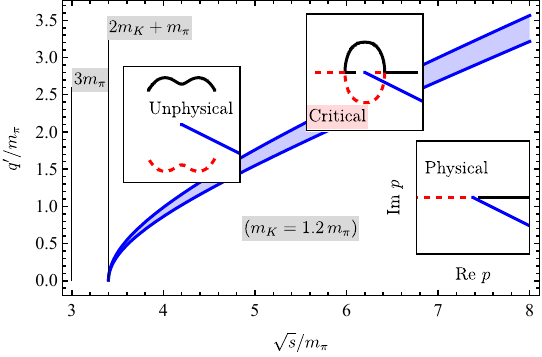}
\end{center}
\caption{Different kinematic regions requiring different methods to analytically continue the spectator momentum to the real axis. The insets show the corresponding locations of three-body cuts according to Eq.~\eqref{eq: singularity3B}. The shown regions correspond to kaon exchange as, e.g., in $KK^*\to \pi(K\bar K)_S$, with modified kaon mass to fit both thresholds in the picture.
}
\label{fig:regions}
\end{figure}

In Fig.~\ref{fig:illustration}, the black (red) dashed  lines indicate the cuts $p_+$ ($p_-$) at fixed $q'$ and $\sqrt{s}$, for $x\in[-1,1]$.
We define the ``critical region'' as the interval for $q'$ in which the cuts form a closed shape as indicated in Fig.~\ref{fig:illustration}. Obviously, one cannot integrate along $\Gamma:=\Gamma_I\cup\Gamma_{II}$ without taking precaution as the path always crosses a cut. The critical region lies at the upper end of the physical region for $q'$. Above and below the critical region, the shapes formed by the cuts are open and analytic continuation to real momenta is simple. For an overview of the different regions, see Fig.~\ref{fig:regions}.
For the unproblematic part of the physical region, ranging from $q'=0$ to the lower limit of the blue shaded regaion, the cuts fully lie on the real $p$-axis, indicating regions in which all three particles are on-shell. The critical (still, physical) region, indicated in blue, exhibits the discussed closed-cut shape. For the unphysical region beyond the critical region, the cuts have fully moved into the complex plane. See also Ref.~\cite{Sadasivan:2021emk} for a similar discussion.

The lower bound of the critical region is determined by $P=0$ where $P:=p_-(m_1,m_3,m_u,-1)$, which is when the end point of the red dashed line is at the origin, see lower right insert of Fig.~\eqref{fig:regions}. We call the branch point $P$ ``triangle point''. 
The condition $P=0$ leads to 
\begin{align}
q'_\text{min}=q_\text{cm}(\sqrt{s}-m_1,m_3,m_u)\ ,
\quad
q_\text{cm}(\sqrt{s},m_a,m_b)=\frac{\sqrt{(s-(m_a-m_b)^2)(s-(m_a+m_b)^2)}}{2\sqrt{s}} \ .
\label{eq: qpmin}
\end{align}
One can interpret this as the outgoing spectator momentum for an incoming isobar/spectator at rest.
The upper bound of the critical region occurs when $P$ leaves the real $p$-axis, i.e., at the end of the physical region (see Fig.~\ref{fig:regions}). 
This occurs when the argument of the square root of Eq.~\eqref{eq: singularity3B} vanishes, again in the kinematics $x=-1$, leading to
\begin{align}
q'_\text{max}=q_\text{cm}(\sqrt{s},m_1+m_u,m_3) \ .
\label{eq:qpmx}
\end{align}
As expected, this momentum corresponds to the smallest possible value of the outgoing isobar mass, $\sqrt{\sigma}_3=m_1+m_u$.
For three pions, one recovers a well-known expression~\cite{Sadasivan:2021emk},
\begin{equation}
    q^{\prime 2}_\text{max}=\frac{9m_\pi^4-10m^2_\pi s+s^2}{4s} \ .
    \label{eq: ppmax}
\end{equation}

%%%%%%%%%%%%%%%%%%%%%%

\subsubsection{Analytic continuation for \texorpdfstring{$\tilde T$}{T}}
In the following we summarize the method of contour deformation to get from complex to real spectator momenta, by Cahill and Sloan~\cite{Cahill:1971ddy, schmid1974quantum}, that was also  applied recently in Ref.~\cite{Pang:2023jri}, see also Ref.~\cite{Zhang:2024fxy}. We expand the method from its original application to the $T$-matrix to the production amplitude $\tilde \Gamma$ needed in this work.

The partial-wave projected scattering equation~\eqref{eq:TLL} can be solved along a complex momentum contour as discussed there. It is straightforward to obtain the solution for real momenta as long as a smooth deformation between complex and real momenta is possible, by simply performing one more iteration. In the uncritical region the integration can be performed as before:
\begin{align}
    \tilde T_{ji}(s,q',p)=\tilde B_{ji}(s,q',p) + \int_\Gamma \frac{dl\,l^2}{(2\pi)^3\,2E_l} \tilde B_{jk}(s,q',l)\tilde \tau_{k}(\sigma_l)\tilde T_{ki}(s,l,p) 
    \label{eq:tildeTnoprob}
\end{align}
with $q'\in [0,q'_\text{min}]$ in the physical, uncritical region, and complex $p,l\in\Gamma$. We have set here and in the following $\tilde C=0$ to operate with shorter expressions. This is not a problem because $\tilde C$ has no cuts in the physical region.

To determine $\tilde T$ for real, outgoing spectator momentum $q'$ in the critical region, $q'_\text{min}<q'<q'_\text{max}$, the smooth deformation from the complex contour $\Gamma$ to  real $q'$ implies the path shown in blue in Fig~\ref{fig:illustration}: first, one integrates $p$ from $0$ to $P(q')$ and back to $Q=0$ changing from first to second Riemann sheet at $P$. This cancels the logarithmic singularity in Re~$\tilde B$ at the branch point $P$, so the integral is numerically stable. The necessary input for this step, $\tilde T(s,q',p)$ with (real) $q'\in [0,P]$ is obtained from Eq.~\eqref{eq:tildeTnoprob} which is possible because $P<q'_\text{min}$ (at least in the one-channel case).

Subsequently, one integrates along $\Gamma$ from 0 to $R$, but on the second sheet, indicated as $\Gamma_{II}$. At $R$, the integration connects analytically back to the first sheet for the rest of the integration along $\Gamma_I$. Altogether, and suppressing channel indices, one has
\begin{multline}
    \tilde T(s,q',p)=\tilde B^I(s,q',p) 
    - 
    \int_0^{P(q')} \frac{dq''\,(q'')^2}{(2\pi)^3 2E_{q''}}\, \text{Im}\,\tilde B^{II}(s,q',q'')\,\tilde\tau(\sigma(q''))\tilde T(s,q'',p)
    \\
    +
    \int_{\Gamma_{II}} \frac{dl\,l^2}{(2\pi)^3\,2E_l} \tilde B^{II}(s,q',l)\tilde\tau(\sigma(l))\tilde T(s,l,p) 
    + 
    \int_{\Gamma_{I}} \frac{dl\,l^2}{(2\pi)^3\,2E_l} \tilde B^I(s,q',l)\tilde \tau(\sigma(l))\tilde T(s,l,p)
       \ .
    \label{eq:tcond}
\end{multline}
The term $-\text{Im }\tilde B^{II}$ comes from the fact that Im~$\tilde B^I=0$ on its first sheet, and $\text{Re } \tilde B^I=\text{Re } \tilde B^{II}$ along the integration path between 0 to $P$. Consequently, only the integration over the imaginary part of $\tilde B$ on the second sheet contributes, from $P$ back to $0$, which explains the minus sign. Furthermore, 
\begin{align}
   \tilde B^{II}=\tilde B^I+2 d \ ,
\end{align}
where the discontinuity $d$ is the analytic function that equals the imaginary part of $\tilde B^I$ for $q''>P$ along the horizontal red dashed line in Fig.~\ref{fig:illustration},
\begin{align}
\text{Im }\tilde B^I=d  \quad\text{for}\quad q''>P \ .
\end{align}
Note that away from the real axis, these two expressions do \emph{not} coincide. 
As Im~$\tilde B^I=0$ for $0<q''<P$, we can replace Im~$\tilde B^{II}\to 2d(s,q',q'')$ in Eq.~\eqref{eq:tcond}. Expressions for $d$ are given in section~\ref{sec:impartB}.

\subsubsection{Analytic continuation for the production amplitude}\label{sec: Ana Con for Gammatilde}

For complex outgoing momentum $p'$ and channel $j$, the production amplitude is obtained from the $\tilde T$-matrix by integrating over all incoming momenta and summing over all incoming channels $i$ (see Fig.~\ref{fig:ccillu}),
\begin{align}
    \tilde\Gamma_j (s,p')=\int_{\Gamma}\frac{\diff p\,p^2}{(2\pi)^3\,2E_p}\,\tilde T_{ji}(s,p',p)\tilde\tau_i(\sigma(p))D_i(s,p) \ ,
    \label{eq:normalgamma}
\end{align}
where $D_i(s,p)$ is a real-valued momentum and energy-dependent production process that can be fit to data. It should obey minimal constraints like the correct centrifugal barrier as $p\to 0$. Optionally, one can introduce Blatt-Weisskopf factors $b_L$, given in Appendix 3 of Ref.~\cite{Mai:2021vsw}, although they can introduce potentially unwanted singularities for complex momenta,
\begin{align}
    D_i(s,p)= D_{fi}(s,p)\, b_{L(i)}(\lambda\, p) \ ,
    \label{eq:D}
\end{align}
with $L(i)$ being the orbital angular momentum of channel $i$.
Here, $D_{fL}$ can have additional dependence on spectator momentum and energy; chosen values are reported in Sec.~\ref{sec:results}. The barrier factors $b_L$ scale with $q^{L(i)}$ at low momenta, but then smoothly approach one at higher momenta. The onset of this suppression is parameterized with a factor $\lambda$~\cite{Mai:2021vsw}. Throughout this study, we chose $\lambda=0.5m_\pi^{-1}$. The full production process also contains a final isobar and its decay, as well as a disconnected piece as discussed in Sec.~\ref{sec:fullprod}. 

Continuing the discussion of analytic continuation of $\tilde \Gamma$ we begin with the non-critical region, because that serves as input for the critical region. Accordingly,  for the subset $0<q'<P<q'_\text{min}$ and with
\begin{align}
\gamma_{j}(s,p')=\int_{\Gamma}\frac{\diff p\,p^2}{(2\pi)^3\,2E_p}\,\tilde B_{ji}(s,p',p)\tilde \tau_i(\sigma(p))D_i(s,p)
\end{align}
one obtains from Eq.~\eqref{eq:tcond} by attaching one more isobar-spectator propagation ($\tilde\tau D$) to all terms:
\begin{multline}
    \tilde \Gamma_{j} (s,q')=\gamma_{j}(s,q')
        -2 
    \int_0^{P(q')} \frac{dq''\,(q'')^2}{(2\pi)^3 2E_{q''}}\, d_{ji}(s,q',q'')\,\tilde\tau_i(\sigma(q''))\,\tilde\Gamma_i(s,q'')
    \\
    +
    \int_{\Gamma_{II}} \frac{dl\,l^2}{(2\pi)^3\,2E_l} \tilde B_{ji}^{II}(s,q',l)\tilde\tau_i(\sigma(l))\,\tilde\Gamma_i(s,l) 
    + 
    \int_{\Gamma_{I}} \frac{dl\,l^2}{(2\pi)^3\,2E_l} \tilde B_{ji}^I(s,q',l)\tilde\tau_i(\sigma(l))\,\tilde\Gamma_i(s,l)
    \ ,
    \label{eq:gcond}
\end{multline}
for $q'$ in the critical region, with input $\tilde \Gamma_i(s,q'')$ on the right-hand side given by Eq.~\eqref{eq:normalgamma}. The only remaining issue is the evaluation of $\gamma$ which is plagued with the same problem of contour deformation as the other terms. The straightforward solution is to chose the same complex path for its evaluation as for the terms containing the $T$-matrix, immediately leading to
\begin{multline}
    \tilde \Gamma_{j} (s,q')=
        -2 
    \int_0^{P(q')} \frac{dq''\,(q'')^2}{(2\pi)^3 2E_{q''}}\, d_{ji}(s,q',q'')\,\tilde\tau_i(\sigma(q''))\left(\tilde\Gamma_i(s,q'')+D_i(s,q'')\right)
    \\
    +
    \int_{\Gamma} \frac{dl\,l^2}{(2\pi)^3\,2E_l} \tilde B_{ji}^{II\to I}(s,q',l)\tilde \tau_i(\sigma(l))\left(\tilde\Gamma_i(s,l) 
    +D_i(s,l)\right)
\ ,
    \label{eq:fancygamma}
\end{multline}
where, for brevity, the analytic function $\tilde B^{II\to I}$ indicates the $\tilde B$ defined on sheet II for $\Gamma_{II}$ and on sheet I for  $\Gamma_I$.

If one plans to re-introduce the contact term $\tilde C$ that was dropped in Eq.~\eqref{eq:tildeTnoprob}, the changes are obvious: On the one hand, the $\tilde T$-matrix entering $\tilde \Gamma_i$ in the noncritical region of Eq.~\eqref{eq:normalgamma} contains now $\tilde C$ according to Eq.~\eqref{eq:TLL}. On the other hand, one substitutes $\tilde B^{II\to I}\to \tilde B^{II\to I}+\tilde C$ in Eq.~\eqref{eq:fancygamma}.

%%%%%%%%%%%%%%%%%%%%%%%%%%%%%%%

\subsubsection{The imaginary part of \texorpdfstring{$\tilde B$}{B} and numerical issues}
\label{sec:impartB}
The imaginary part $d$ along the real $q''$ axis in Eq.~\eqref{eq:fancygamma} is given by
\begin{align}
d_{ji}(s,q',q'')=\frac{N_{ji}(x_0)\,d_f}{2E_{q'+q''}(x_0)}
\quad\text{with }\quad
d_f=\text{Im}\int\limits_{-1}^1 dx\,\frac{1}{
f(\bm{q}'',\bm{q}')}=-\frac{i\pi(\sqrt{s}-E_{q'}-E_{q''})}{q'q''}
\label{eq:explim}
\end{align}
with $f$ from Eq.~\eqref{eq:ffunc}, i.e.,
 $d_f$ is given by the imaginary part of $\tilde B$-term with unity in the numerator (and is, therefore, also channel dependent). In Eq.~\eqref{eq:explim}, the quantity
\begin{align}
x_0=\frac{m_1^2+m_3^2-m_u^2+2E_{q''}E_{q'}-2\sqrt{s}(E_{q''} +E_{q'})+s }{2q'q''} \ ,
\end{align}
with $E_{q''}=\sqrt{m_1^2+q''^2}, E_{q'}=\sqrt{m_3^2+q'^2}$,
is the position of the $\tilde B$ singularity in $x=\cos\theta$, given by $f(\bm{q}'',\bm{q}')=0$. As an example for the numerator of the $\tilde B$-term, consider a transition from a channel with vector meson $i$ to another channel with vector meson, $j$. Following Eq.~\eqref{btilde} and the partial wave decomposition~\eqref{eq:b pw projection}, one obtains
\begin{align}
N_{ji}(x)=2\pi\,  \tilde I_f U_{L(j)\lambda'}v^*_{\lambda'}(P-p-p',p)v_{\lambda}(P-p-p',p')d^J_{\lambda'\lambda}(x)   U_{\lambda L(i)} \ ,
\end{align}
including the transformation from helicity to JLS basis with $U$ from Eq.~\eqref{Umatrix}. Other numerators, as given in Eq.~\eqref{eq:v}, are simpler and evaluated accordingly. 

One should note that the partial-wave integrals of Eq.~\eqref{eq:b pw projection} for $\tilde B$ in Eq.~\eqref{eq:fancygamma} need to be performed for complex incoming and real outgoing spectator momenta. One possibility to avoid the first-order singularity in $x$, that sometimes occurs, is to deform the $x$-integration in between its endpoints at $x=\pm 1$. We found that a smooth, complex path that maximally extends up to $x=-0.2\,i$ into the complex plane (e.g., along a parabola) greatly speeds up the integration requiring fewer Gauss points.

%%%%%%%%%%%%%%%%%%%%%%%%%%%%%%%%%%%

\subsection{Direct inversion of the integral equation}
\label{sec:direct}
Another method of solving the integral equation is also discussed in Ref.~\cite{schmid1974quantum}, see also the original reference~\cite{Sohre:1971hy}.
We refer to it as direct inversion (DI) as compared to the contour deformation method (CD) discussed before. There, we first discussed the continuation for the $\tilde T$ matrix and then for $\tilde\Gamma$. Here, we consider directly the scattering equation for the sum of $\tilde\Gamma$ and the disconnected part, $\tilde\Gamma^T=\tilde\Gamma+D$, that reads in operator notation $\tilde\Gamma^T=D+\tilde B\,\tilde \tau\,\tilde\Gamma^T$ or, explicitly,
\begin{align}
\tilde\Gamma^T_j(s,q')=D_j(s,q')+\int_0^\Lambda \frac{dq\,q^2}{(2\pi)^3 2E_{q}}\,
\tilde B{ji}(s,q',q)
\,
\tilde\tau_i(\sigma(q))
\,
\tilde\Gamma^T_i(s,q) \ ,
\label{dirprod}
\end{align}
with the integration taken along the real axis and $E_q$ is the spectator energy as before. Note that the full production process requires $\tilde \Gamma^T$ to be multiplied by one more isobar-spectator propagator, see Sec.~\ref{sec:fullprod}.
The problem of moving three-body singularities can be shifted from the integral equation to individual integrals that are easier to treat numerically as explained in the following.

Following Ref.~\cite{schmid1974quantum}, $\tilde \Gamma^T$ is suitably discretized in $q$, assuming that the solution can be well approximated by an interpolating polynomial connecting the discrete $\tilde \Gamma^T(q_i)$ (channel index and index $s$ omitted),
\begin{align}
  \tilde \Gamma^T(q)\approx
  \sum_{i=1}^N \tilde \Gamma^T(q_i)\,H_i(q) \ ,
  \label{ansatzDI}
\end{align}
with Lagrange polynomials 
\begin{align}
    H_i(q)=
\frac{\prod_{j\neq i}^N (q-q_j)}{\prod_{j\neq i}^N(q_i-q_j)} \ .
\end{align}
The $N$ real momenta $q_i$ can be freely chosen, even beyond the physical region as demonstrated in the numerical example below.
Alternatively, one can approximate $\tilde\Gamma^T$ by splines of a given order, which requires a certain care in choosing appropriate grid points~\cite{Matsuyama:2006rp}. 
Inserting the Ansatz~\eqref{ansatzDI} into Eq.~\eqref{dirprod}, 
one obtains 
a set of linear equations that can be easily inverted,
\begin{align}
\tilde \Gamma^T(q_j)=D(q_j) +A_{ji} \tilde \Gamma^T(q_i)   
\end{align}
with
\begin{align}
    A_{ji}=\int_0^\Lambda \frac{dq\,q^2}{(2\pi)^32E_q}\,\tilde B (q_j,q)\, \tilde \tau(\sigma(q))\,H_i(q) \ .
\end{align}
These integrals have logarithmic singularities in $q$ from the endpoints of the angular integrations according to Eq.~\eqref{eq:b pw projection}.
These singularities could be treated similar to Ref.~\cite{Matsuyama:2006rp} where a variable transformation is used to remove them. Alternatively, modern adaptive/iterative integration routines such as provided by Mathematica are sufficiently  effective to directly solve these integrals as the position of the logarithmic singularities is analytically known (see Sec.~\ref{sec:anal}). We also note that the first-order singularities occurring in the partial-wave integrals of Eq.~\eqref{eq:b pw projection} are a problem due to having real spectator momenta. See Sec.~\ref{sec:impartB} where a solution to this problem in terms of a complex angular integration was discussed.

%%%%%%%%%%%%%%%%%%%%%%%%%%%
\begin{figure}[htb]
\begin{center}
\includegraphics[width=0.5\textwidth]{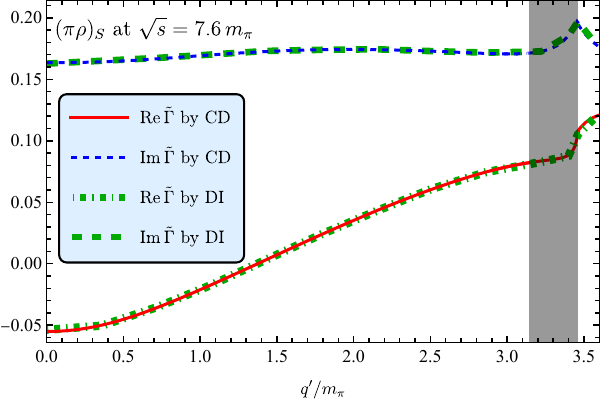}
\end{center}
\caption{Comparison of calculating the production amplitude $\tilde\Gamma$ through contour deformation (CD) or direct inversion (DI). Both methods agree well and exhibit a cusp at the largest physical spectator momentum, at the upper end of the critical region (shaded). For simplicity, the model is here reduced to one channel, $\pi\rho$ in S-wave.}
\label{fig:methods}
\end{figure}

In Fig.~\ref{fig:methods} the CD and DI methods are numerically compared. Both methods agree within small deviations depending on the chosen grid of spectator momenta $q'$ of the DI method. The figure shows the ``critical region'' of the CD method shaded in gray; compare also Fig.~\ref{fig:regions}. At the upper end of the physical region (right edge of the gray area), we observe a cusp. Note that the imaginary part of the amplitude beyond the physical region, $q>q'_\text{max}$, does not vanish. It means that, despite the outgoing state being off-shell, there are still intermediate loops in the rescattering series that do contribute with an imaginary part.

Comparing the CD and DI methods, the DI is clearly simpler although solving integrals with logarithmic singularities still poses a certain challenge. One advantage of the DI method lies in the fact that these integrals can be pre-computed. An advantage of the CD method might be the possibility to use the solution on the complex contour not only to calculate the amplitude with real momenta, but also to search for resonance poles on unphysical Riemann sheets~\cite{Sadasivan:2021emk}. A certain disadvantage of the CD method lies in the fact that one needs to explicitly know the channel-dependent locations of the points $P$ and $R$, see Fig.~\ref{fig:illustration}. This is always possible but might be cumbersome to implement. Both methods will have to be adapted once the propagation $\tilde\tau$ itself has singularities on the real axis, as it occurs for channels with two stable particles. An example is the coupled $\pi N$, $\pi\pi N$ system~\cite{Ronchen:2012eg}. 
%%%%%%%%%%%%%%%%%%%%%%%%%%%%%%%%%%%

\subsection{Production amplitude}
\label{sec:fullprod}
So far, we calculated the connected, amputated rescattering term $\tilde \Gamma$ in Eq.~\eqref{eq:gcond}. The production of three real particles in the final state requires the attachment of one final isobar and its decay vertex,
\begin{align}
    \breve\Gamma_{j}(s,q')=\breve v_{j}(\sigma(q'))\tilde\tau_{j}(\sigma(q'))\left[\tilde\Gamma_{j}(s,q')+D_{j}(s,q')\right]
    \label{eq:gammabrev}
\end{align}
with $D$ from Eq.~\eqref{eq:D} and $\tilde \Gamma+D$ called $\tilde \Gamma^T$ in Sec.~\ref{sec:direct}.
In general, the final decay vertices are angle dependent. For the S-wave amplitudes, they are simply given as $\breve v_i=1$ because $\tilde \tau$ corresponds already to the full amplitude in plane-wave basis. We define the final decay vertex of the $\rho$ and $K^*$ as the plane-wave 2-body decay vertex, with the angular part (Wigner-$D$ functions) removed, 
\begin{align}
    \breve v_j=\sqrt{\frac{2\ell+1}{4\pi}}\, \tilde v_{V_jMM'} \ ,
\end{align}
with $\ell=1$ and $\tilde v$ from Eq.~\eqref{eq:vpw} and the three-momenta of the $M$ (pseudoscalar mesons) defined in the vector meson rest frame through $\sigma(q')$.
The construction of a proper production amplitude requires the possible symmetrization of $\breve\Gamma$ for identical particles in the outgoing state (e.g., two $\pi^-$) and the attachment of the angular structure~\cite{JPAC:2018zwp,Sadasivan:2020syi, Sadasivan:2021emk}. As this manuscript is not aimed at data analysis, we stay with the production amplitude in the form of Eq.~\eqref{eq:gammabrev} and study its properties in the following.

%%%%%%%%%%%%%%%%%%%%%%%%%%%%%%%%%%%%%%%%%%%%%%%%%%%%%%%
%%%%%%%%%%%%%%%%%%%%%%%%%%%%%%%%%%%%%%%%%%%%%%%%%%%%%%%
\section{Numerical Results}
\label{sec:results}
%%%%%%%%%%%%%%%%%%%%%%%%%%%%%%%%%%%%%%%%%%%%%%%%%%%%%%%
%%%%%%%%%%%%%%%%%%%%%%%%%%%%%%%%%%%%%%%%%%%%%%%%%%%%%%%
To demonstrate the feasibility of the present approach, a few numerical results are presented in the following. 
We focus on three-pion production, so 
the three-body energy is set to some value between the two thresholds, chosen as $\sqrt{s}=7\,m_\pi$. The numerical methods presented in the previous section work also above the $\pi KK$ threshold.

The free parameters of the amplitude have to be fixed. In particular, we chose all contact terms from Eq.~\eqref{eq:TLL} $\tilde C=0$. For the production vertices $D_f$ from Eq.~\eqref{eq:D}, we chose: $D_{f}=1\,m_\pi$ for channels with vector isobars, $\ell(i)=1$, and $D_{f}=1$ for scalar isobars $\ell(i)=0$ to make the
 overall production amplitude $\breve\Gamma$ from Eq.~\eqref{eq:gammabrev} unitless.
 
\begin{figure}[htb]
  \centering
\setlength{\unitlength}{1\textwidth}
  \begin{picture}(1,0.52)
%\includegraphics[width=0.9
%\textwidth]{Gammatilde(pisigmaP isolated).PDF}
\put(0.05,0){\includegraphics[width=0.9\textwidth]{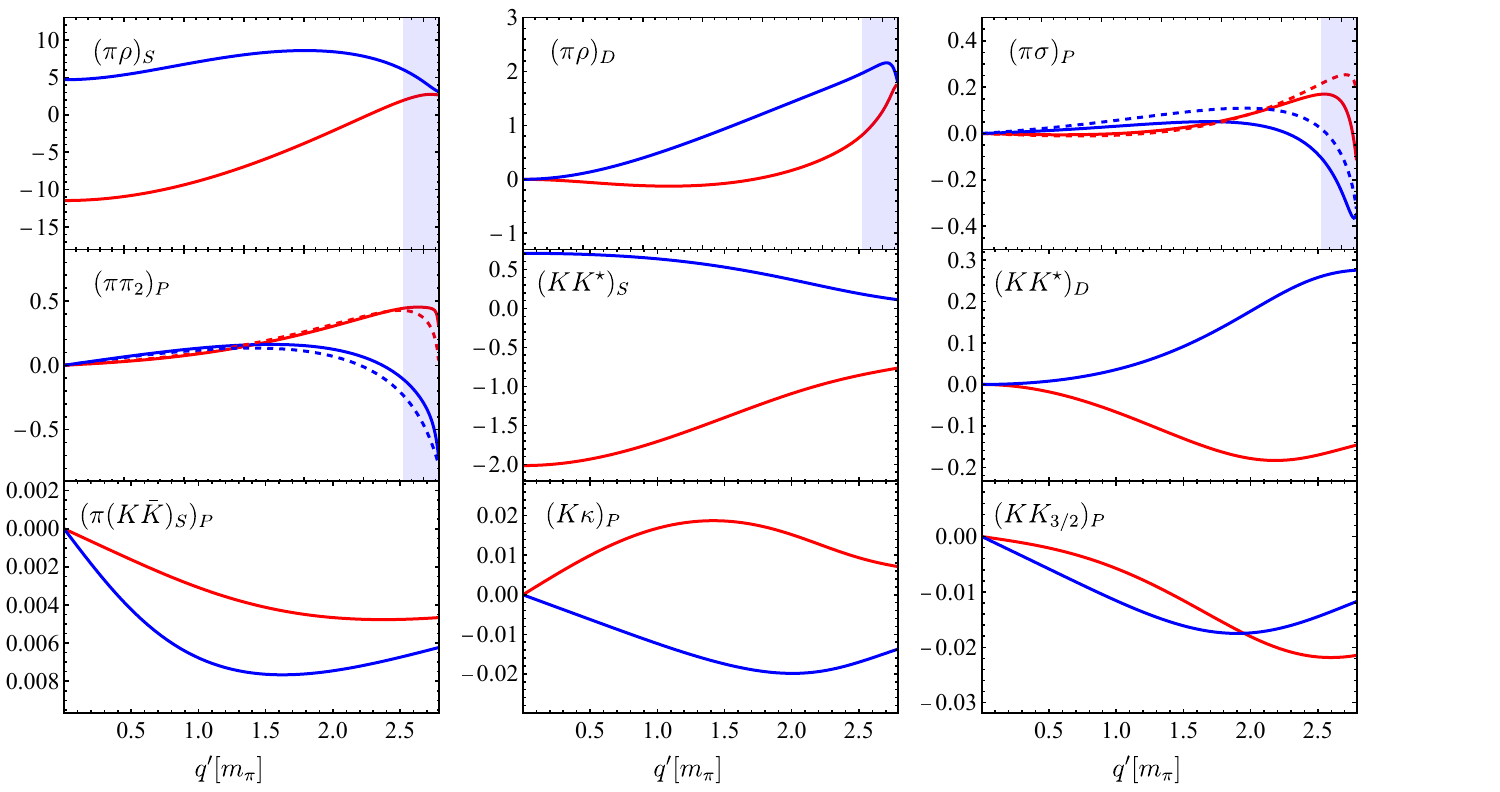}}
\put(0.02,0.28){
$\tilde\Gamma$}
\end{picture}
\caption{Re-scattering part $\tilde{\Gamma}$ for the 9-channel model of all partial waves at a 3-body energy of $\sqrt{s}=7\,m_\pi$. 
The outgoing channel names according to Table~\ref{tab:channels} are indicated.
The units of $\tilde \Gamma$ are the same as for $D$, i.e., the y-axes show $\tilde\Gamma/m_\pi$ if the outgoing isobar is a vector ($\ell=1$) and unitless $\tilde \Gamma$ for outgoing scalar isobars ($\ell=0$). The red and blue curves represent the real and imaginary parts, respectively. The dashed curves for the $\pi\sigma$ and $\pi\pi_2$ final states represent the model with $\pi\rho$  channels switched off. The blue shaded areas indicate ``critical regions'' as defined in Sec.~\ref{sec:critical}.
}
\label{fig:Gammatilde-9channel}
\end{figure}

Fig.~\ref{fig:Gammatilde-9channel} shows the real and imaginary parts of the re-scattering term $ \tilde{\Gamma}$ of all nine channels, using the contour deformation method described in Sec.~\ref{sec: Ana Con for Gammatilde}.  At the chosen three-body energy, the ``critical region'' (blue shaded area) only appears in the channels with three-pion final state, because the outgoing channels with strangeness are sub-threshold.  
It is obvious that the CD method from Sec.~\ref{sec:anal} provides a smooth transition from the non-critical to the critical region when evaluating production amplitudes at the real, physical momenta. 
The $\tilde\Gamma$ for the $\pi\sigma$ and $\pi\pi_2$ channels exhibit rapid phase motion in the critical region. It is found that the this is partly a result of coupled channel effects from the large $\pi\rho$ channel. This can be seen by switching off the $\pi\rho$ channel, resulting in the dashed lines in Fig.~\ref{fig:Gammatilde-9channel}. 

\begin{figure}[htb]
  \centering
\setlength{\unitlength}{1\textwidth}
  \begin{picture}(1,0.4)
\put(0.2,0){
\includegraphics[width=0.6\textwidth]{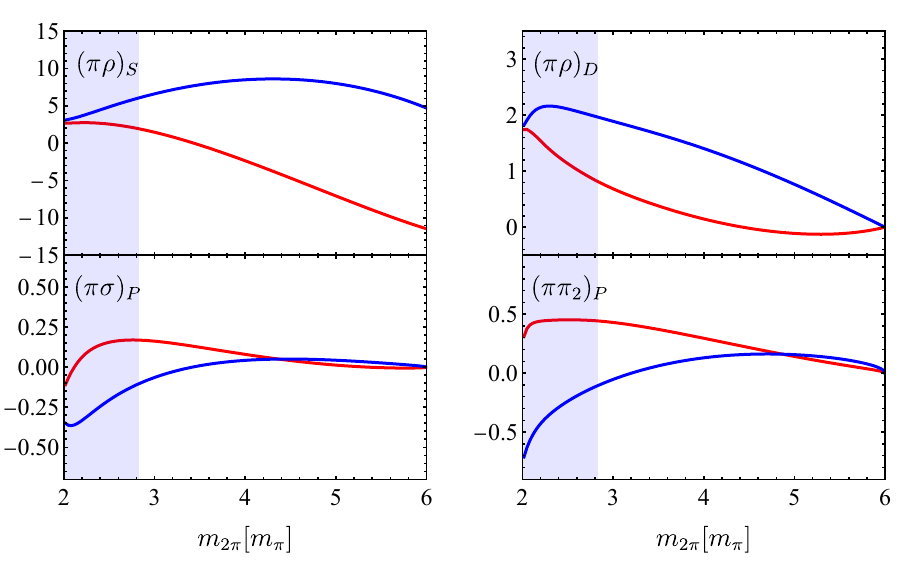}}
\put(0.17,0.21){
$\tilde\Gamma$}
\end{picture}
\caption{Rescattering term $\tilde{\Gamma}$ as function of the isobar invariant mass, in units of $m_\pi$ for vector isobars and unitless for scalar isobars. The red and blue curves represent the real and imaginary parts. 
%See Fig.~\ref{fig:Gammatilde-9channel} for more information.
}
\label{fig:Gammatilde 3pion (isobar)}
\end{figure}

Fig.~\ref{fig:Gammatilde 3pion (isobar)} shows the  same $\tilde\Gamma$ of the three-pion final states   as a function of the $\pi\pi$ invariant mass. In this representation one observes substantial phase motion for the $(\pi\rho)_D$ final state. All $\tilde\Gamma$ with $L>0$ vanish at the largest invariant mass due to their centrifugal barriers, as expected. 
We also observe that the $(\pi\rho)_S$ channel is dominant despite the input $D_{fi}$ being of comparable size for all channels. Clearly, all other channels are much smaller due to their centrifugal barriers. There is also an enhancement of outgoing channels with vector isobars in $\tilde\Gamma$, due to the structure of the $\tilde B$-term itself. According to Eq.~\eqref{btilde}, there is one $VMM$ vertex too many due to the splitting of the $\ell=1$ two-body amplitude across $\tilde B$ and $\tilde\tau$. This is only the case for the (unphysical) quantity $\tilde\Gamma$ and not for the full amplitude  $\breve\Gamma$, that is physical up to inclusion of angles and symmetrization of final states.

\begin{figure}[htb]
\begin{center}
\includegraphics[width=0.45\textwidth]{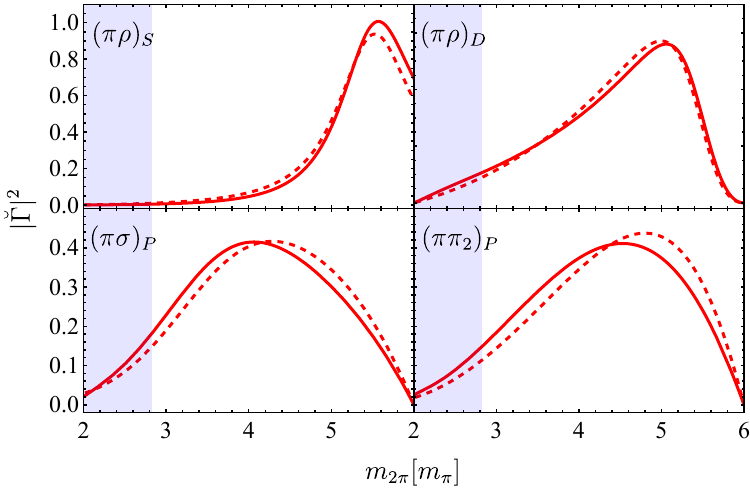}
\includegraphics[width=0.45\textwidth]{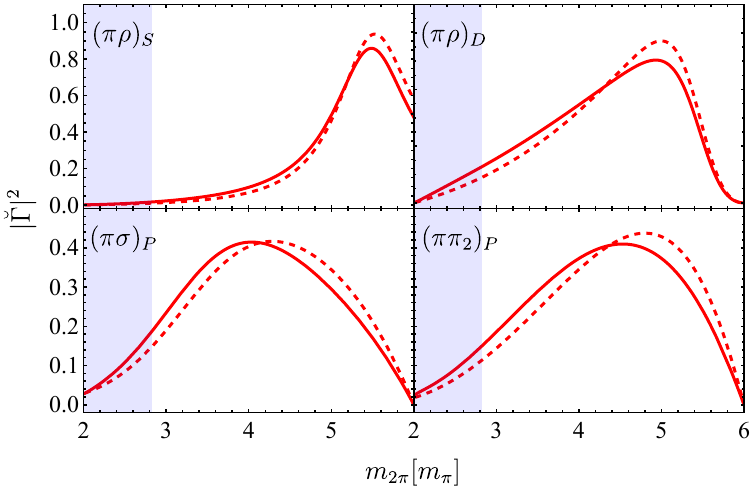}
\end{center}
\caption{Production amplitude $|\breve{\Gamma}|^2$ (unitless for all channels) in terms of the isobar invariant mass. The left figure shows the result with the standard choice for the production vertex $D$, while the right figure corresponds to modified input.
The red dashed lines show the disconnected contributions that correspond to  tranditional isobar amplitudes/lineshapes, while the red solid lines correspond to the full, three-body unitary amplitude including coupled channels and meson exchanges. The areas under all curves are normalized to one to facilitate the comparison of the lineshapes. The light blue area indicates the critical region.}
\label{fig:normalized production amplitude}
\end{figure}

In Fig.~\ref{fig:normalized production amplitude}, to the left, we show the complete production amplitude $\breve\Gamma$ from Eq.~\eqref{eq:gammabrev} including the disconnected part,  the final isobar propagation, and its decay. Consequently, one observes the $\rho$ and $\sigma$ lineshapes, but not the $f_0(980)$ lineshape, due to the chosen energy being $\sqrt{s}=7\,m_\pi$. 
The dashed lines show the disconnected part alone. This would correspond to the traditional isobar model, in which lineshapes are related to two-body scattering amplitudes only. The solid lines show the full, three-body unitary amplitude $|\breve\Gamma|^2$, including coupled channels and rescattering.  
There is a noticeable modifications of the lineshapes by these effects, at some invariant masses. We observe a distortion of the $\sigma$ meson towards smaller invariant masses, and a certain accumulation of strength at the two-pion threshold for the channels with S-wave isobars, $(\pi\sigma)_P$ and $(\pi\pi_2)_P$. 

If one chooses the input for the production vertex $D_{f}$ differently, one obtains different line shapes. As an example, we show in Fig.~\ref{fig:normalized production amplitude} to the right the outcome for
$D_{f}=3\,m_\pi$ for the $(\pi\rho)_S$ and $(KK^*)_S$ channels, $D_{f}=2\,m_\pi$  for the $(\pi\rho)_D$ and $(KK^*)_D$ channels, and $D_{f}=0.5$ for the scalar isobars.  Of course, there is additional freedom in the amplitude that will further modify the lineshapes. This refers to the isobar-spectator contact terms $\tilde C$ in Eq.~\eqref{eq:TLL}. These terms encode three-body resonance dynamics~\cite{Sadasivan:2021emk} and other QCD effects, and they have all been set to zero in this study for simplicity.

%%%%%%%%%%%%%%%%%%%%%%%%%%%%%%%%%%%%%%%%%%%%%%%%%%%%%%

\section{Summary and Conclusions}
\label{sec: conclusion}
A unitary three-body amplitude for pions and kaons is derived, expanding upon previous work that only included the $\pi\rho$ channel which is dominant in the $a_1(1260)$ meson. In this study, two-body subsystems (``isobars'') for all isospins up to P-waves are included, with both resonant and repulsive properties, namely 
the $S-$wave isobars
$\sigma,\,\kappa$, isospin-2 $\pi\pi$ (named $\pi_2$), isospin-3/2 $K\pi$ (named $K_{3/2}$), and one more P-wave isobar in form of the $K^\star(892)$.
In addition, the $\sigma$ isobar allows for $\pi\pi\leftrightarrow K\bar K$ transitions. 
The two-body unitary coupled-channel dynamics is parameterized using simple chirally inspired interactions, and the vector isobars rely on a separable 
ansatz with explicit resonances. The calculation of isospin coefficients is performed in two different ways to check consistency.

The amplitude is evaluated for complex momenta and then analytically continued to real, physical, spectator momenta, for the example of $a_1$ quantum numbers. Identical results are obtained by a direct inversion of the amplitude; advantages and disadvantages of both methods are discussed. The internal consistency of the contour deformation method is 
verified
numerically in Sec.~\ref{sec:results}.

In future work, the nine-channel production amplitude can be used to analyze semileptonic $\tau$ decays and related reactions by tuning real-valued, channel-, momentum-, and energy-dependent contact terms $\tilde C$ and production terms $D$ to data. For this purpose, obvious extensions concern the introduction of $\eta$ mesons and higher-spin isobars.

Also, the properties of the amplitude compared to the traditional isobar picture will be further explored, especially regarding the modification of isobar lineshapes through 3-body rescattering effects. For this, the energy range has to be extended to make sure to capture the full COMPASS kinematic range of their lineshapes. The amplitude is also prepared to study the $a_1(1420)$ triangle singularity as a coupled-channel effect. Furthermore, extensions to other quantum numbers will allow to study dynamical resonance generation, exotic channels, and extensions to finite volume for the analysis of future lattice QCD data. 

%%%%%%%%%%%%%%%%%%%%%%%%%%%%%%%%%%%%%%%%%%%%%%%%%%%%%%
\begin{acknowledgments}
The authors thank Kanchan Khemchandani, Alberto Martinez Torres, and Toru Sato for discussions.
 This work was supported by the U.S. Department of Energy contract DE-AC05-06OR23177, under which Jefferson Science Associates, LLC operates Jefferson Lab. Also 
 by the 
 U.S. Department of Energy ``ExoHad'' Topical Collaboration, contract DE- SC0023598, the 
U.S. Department of Energy Grants: ~DE-SC0016582, 
 DE-FG02-87ER40365, and the  National Science Foundation (NSF) Grant No. 2310036.
% The work of YF, MD, and MM is supported by the National Science Foundation (NSF) Grant No. 2310036. 
%This work contributes to the aims of the U.S. Department of Energy ExoHad Topical Collaboration, contract DE- SC0023598. 
%This work is supported by the U.S. Department of Energy ``ExoHad'' Topical Collaboration, contract DE- SC0023598.
%This work is also supported by the U.S. Department of Energy grant DE-SC0016582.
%and Office of Science, Office of Nuclear Physics under contract DE-AC05-06OR23177. 
%%
MM acknowledges funding by the Deutsche Forschungsgemeinschaft (DFG, German Research Foundation) through the Sino-German Collaborative Research Center TRR110 “Symmetries and the Emergence of Structure in QCD” (DFG Project ID 196253076 - TRR 110), as well as the Heisenberg Programme (project number: 532635001).
R. M. acknowledges support from the CIDEGENT program
with Ref. CIDEGENT/2019/015, the Spanish Ministerio de
Economia y Competitividad and European Union (NextGenerationEU/PRTR) by the grant with Ref. CNS2022-136146. This work is also partly supported by the Spanish Ministerio de Economia y Competitividad (MINECO) and European FEDER funds under Contracts No. FIS2017-84038-C2-1-P B, PID2020-
112777GB-I00, and by Generalitat Valenciana under contract PROMETEO/2020/023.
\end{acknowledgments}

%%%%%%%%%%%%%%%%%%%%%%%%%%%%%%%

\appendix
\section{Connection to Lagrange formalism}
\label{sec:lagrangians}
It is useful to double check the isospin coefficients of at least some of the transitions derived in Sec.~\ref{sec:transitions}. For this we consider a Lagrangian describing vector mesons coupling to pseudoscalar mesons, 
\begin{align}
    {\cal L}=\frac{ig}{4}
    \text{Tr}\left( V_\mu\left[\partial^\mu \Phi,\Phi\right]\right) \ ,
\end{align}
where $g=\frac{m_\rho}{\sqrt{2}f_\pi}$ and
\begin{align}
    V_\mu=
    \begin{pmatrix}
        \rho^0_\mu+\omega_\mu & \sqrt{2}\rho_\mu^+ & \sqrt{2}K_\mu^{*+}\\ 
        \sqrt{2}\rho_\mu^- & -\rho^0_\mu+\omega_\mu &\sqrt{2}K_\mu^{*0}\\
        \sqrt{2}K_\mu^{*-} & \sqrt{2}\bar K_\mu^{*0} &\sqrt{2}\phi_\mu
    \end{pmatrix},\quad
    \Phi=\begin{pmatrix}
    \sqrt{\frac{2}{3}}\eta+\frac{\eta'}{\sqrt{3}}+\pi^0 & \sqrt{2}\pi^+ &\sqrt{2} K^+ \\
    \sqrt{2}\pi^- & \sqrt{\frac{2}{3}}\eta+\frac{\eta'}{\sqrt{3}}-\pi^0 &\sqrt{2} K^0 \\
    \sqrt{2} K^- & \sqrt{2} \bar K^0 & -\sqrt{\frac{2}{3}}\eta+\frac{2\eta'}{\sqrt{3}} 
    \end{pmatrix} \ ,
\end{align}
assuming $\eta-\eta'$ and $\rho-\omega$ ideal mixing with $\eta_8=\frac{2\sqrt{2}}{3}\eta-\frac{1}{3}\eta'$ and $\phi=\omega_1/\sqrt{3}-\omega_8\sqrt{2/3}$, $\omega=\omega_1\sqrt{2/3}+\omega_8/\sqrt{3}$.
For an exchange diagram using $i{\cal H}=-i{\cal L}$ we obtain the vertex for the u-channel exchange:
\begin{align}
    -iV_u=(-it_1)(i\Delta_u)(-it_2) \ ,
\end{align}
with incoming and outgoing $VMM'$ vertices $t_i$ and meson exchange propagator $\Delta_u$, where $M$ ($V$) stands for a pseudoscalar (vector) meson, not to be confused with vertex $V_u$. Replacing the vector fields by polarization vectors, we obtain the following transitions in the isospin basis:
\begin{align}
    V_u=\frac{I_{\cal L} V^*_{\lambda'}(p,P-p'-p)V_{\lambda}(P-p'-p,p')}{u-m_\pi^2},\quad V_\lambda(q_2,q_3)=-ig\epsilon_\lambda^\mu(q_2-q_3)_\mu \ ,
    \label{tvec}
\end{align}
where $p(p')$ and $\lambda(\lambda')$ are the incoming (outgoing) spectator momenta and isobar helicities, according to Fig.~\ref{fig:exlabels}.
From the Lagrangian we obtain the following Lagrangian-based isospin factors $I_{\cal L}$ for definite total isospin $I$: 
\begin{align}
\begin{tabular}{l|rrr}
\hline\hline
$I$ & 0 & 1 &2  \\ \hline
$I_{\cal L}(\pi\rho\to \pi\rho)$ & $2$ & $-1$ & $-1$  \\
$I_{\cal L}(\pi\rho\to KK^*)$ & $-\nicefrac{\sqrt{3}}{2}$ & $\nicefrac{1}{\sqrt{2}}$ & 0\\
$I_{\cal L}(KK^*\to KK^*)$ &  $\nicefrac{3}{4}$ & $\nicefrac{1}{4}$ & 0
\\ \hline\hline
\end{tabular} \ .
\end{align} 
Similarly, one can calculate the isospin factor $I(VMM')$ for the $V\to MM'$ transition with the result quoted below  Eq.~\eqref{eq:tauv}. These factors, multiplied with the Clebsch-Gordan combinations of Eqs.~\eqref{eq:isofac} and \eqref{eq:isofacstrange} should equal the isospin coefficients from the Lagrangian formalism, 
\begin{align}
    I(V'M_i M_i')I(VM_i M_i')\tilde I_F=-I_{\cal L} \ ,
\end{align}
for incoming (outgoing) vector meson $V(V')$, up to a minus sign coming from the exchanged order of arguments in one of the vertices of Eq.~\eqref{btilde} compared to Eq.~\eqref{tvec}. This equality holds, indeed, if one chooses the lower sign in Tables~\ref{tab:I=0 IFtilde} to \ref{tab:I=2 IFtilde}, which is conventional. Note also that in Eq.~\eqref{btilde} we have explicitly canceled the factor $i(-i)$ from $V^*_{\lambda'}V_{\lambda}$. This makes no difference for $VM\to VM$ transitions, but it leads to un-observable phases in some off-diagonal channel transitions of the $T$-matrix. However, if one works with explicit factor of $i$ in the definition of $V$ when using Lagrangians, then the production vertex $D$ should also contain  explicit, channel dependent factors of $i$. This is possible but cumbersome. It is easier to consistently remove the factors of $i$ from the $V$, making the $\tilde B$ term in the JLS basis and the $D$ term real for all channels and real momenta and energies. 

%%%%%%%%%%%%%%%%%%%%%%%%%%%%%%%
%%%%%%%%%%%%%%%%%%%%%%%%%%%%%%%%%%%%%%%%%%%%%%%%%%%%%%%%%
%\bigskip
%%%%%%%%%%%%%%%%%%%%%%%%%%%%%%%%%%%%%%%%%%%%%%%%%%%%%%%%%
%%%%%%%%%%%%%%%%%%%%%%%%%%%%%%%%%%%%%%%%%%%%%%%%%%%%%%%%%
\bibliography{BIB}
%%%%%%%%%%%%%%%%%%%%%%%%%%%%%%%%%%%%%%%%%%%%%%%%%%%%%%%%%

\end{document}